\documentclass[
    amsmath,amssymb,
    aps,twocolumn,
    prc,
    floatfix
]{revtex4-2}

\usepackage{graphicx}
\usepackage{dcolumn}% Align table columns on decimal point
\usepackage{bm}
\usepackage{hyperref}% 

\begin{document}

\preprint{}

\title{Systematic analysis of inner crust composition using the extended Thomas-Fermi approximation with pairing correlations}

\author{Matthew Shelley}
    \email{mges501@york.ac.uk}
\author{Alessandro Pastore}
    \email{alessandro.pastore@york.ac.uk}
\affiliation{
    University of York, Heslington, York, YO10 5DD, United Kingdom, 
}

\date{\today}

\begin{abstract}
We perform a systematic investigation of the chemical composition of the inner crust of a neutron star, using the extended Thomas-Fermi approximation, the Strutinsky integral correction for shell effects, and the BCS approximation for pairing. Fifteen Skyrme functionals were selected, which cover the range of values of important bulk properties of infinite nuclear matter, while also having pure neutron matter (PNM) equation of states (EoS) with varying degrees of stiffness. We find that a functional's low-density PNM EoS is correlated with the number of protons found in the inner crust's nuclear clusters and, in the lower-density region of the inner crust, with the pressure.
\end{abstract}

\maketitle

%%%%%%%%%%%%%%%%%%%%%%%%%%%%%%%%%%%%%%%%%%%%%%%%%%%%%%%%%%%%%%%%%%%%%%%%%%%%%%%%%%%%%%%%%%%%%%%%%%%%

\section{Introduction}\label{sec:intro}

Due to a strong pressure gradient, the matter within a neutron star (NS) organises itself into different layers~\cite{chamel2008physics}. Near the surface of the star lies the \emph{crust}, where neutrons and protons are arranged in nuclear clusters in a crystalline structure, that transitions gradually into a liquid phase~\cite{mantziris2020neutron} approaching the core of the star~\cite{xu2009locating}.

Although the crust only accounts for a small fraction of the mass of the NS, it plays a major role in a variety of astrophysical phenomena, including the r-process~\cite{lattimerDecompressionColdNeutron1977}, short gamma-ray burst precursors caused by resonant shattering~\cite{tsangResonantShatteringNeutron2012}, soft gamma-ray repeater giant flares~\cite{thompson1995soft, strohmayer20062004}, and thermal relaxation in soft x-ray transients. The equation of state (EoS) of the crust is also believed to influence many properties of NS, such as the moment of inertia~\cite{piekarewiczPulsarGlitchesCrust2014} which influences pulsar glitches, transport phenomena within the star~\cite{link1999pulsar, burrows2006neutrino, brown2000nuclear}, the relation between the radius and tidal deformability in low-mass NS~\cite{ferreiraEffectCrustNeutron2020a}, and the value of the second Love number,~$k_2$~\cite{piekarewiczImpactNeutronStar2019, perotRoleCrustTidal2020}. Therefore it is crucial to have a reliable model of the crust of a NS, to achieve accurate descriptions of these phenomena.

As discussed in Ref.\cite{chamel2008physics}, current models predict that the crust has a crystalline structure, comprising neutron-rich nuclei surrounded by ultra-relativistic electrons~\cite{baym1971ground, pastore2020impact, fantina2020crystallization}.
The crust can be further separated into the \emph{inner} and \emph{outer} parts. The \emph{outer} crust is the lower-density part of the crust, where neutrons are found only in bound states. The \emph{inner} crust begins above baryonic densities of $n_b\approx10^{11}$~g\thinspace cm$^{-3}$, where neutrons begin to drip out of the nuclei, forming a gas.

Since the pioneering work of Negele and Vautherin~\cite{negele1973neutron}, many theoretical models have been developed to study the properties of the inner crust. The presence of a neutron gas dramatically changes the properties of the clusters~\cite{avogadro2007quantum}, in contrast to the outer crust, whose nuclei are isolated. Consequently, determining the composition of the inner crust requires a simultaneous treatment of the clusters and the neutron gas. To simplify this task, it is customary to adopt the Wigner-Seitz~(WS) approximation, in which the inner crust is decomposed into independent spheres, named \emph{WS cells}, centered around each cluster, with a radius $R_{\mathrm{WS}}$. Each cell is at $\beta$-equilibrium, with a certain number of protons, $Z$, and the same number of electrons (under the condition of charge-neutrality) spread through the cell. Using this approximation, the values of $Z$ (the \emph{chemical composition}) and $R_{\mathrm{WS}}$ can be determined at a given baryonic density, $n_b$, by minimising the total energy per particle, $e$, of the cell~\cite{pearsonInnerCrustNeutron2012}. This procedure is valid only in the zero-temperature limit, which is applicable to the case of a non-accreting NS.

Since the density of electrons in the crust is essentially uniform~\cite{watanabe2003electron}, it is possible to calculate their contribution to $e$ analytically~\cite{chamel2008physics}. The nuclear contribution is more complex and requires the use of a model. In the literature, several models are used to determine the nuclear binding energy of the system, such as the compressible liquid drop model~\cite{douchin2000inner}, semi-classical models using the Thomas-Fermi approximation~\cite{oyamatsu1994shell, onsiEquationStateStellar1997a, onsi2008semi, pearsonInnerCrustNeutron2012, martinLiquidgasCoexistenceEnergy2015, sharmaUnifiedEquationState2015, limStructureNeutronStar2017}, or the Hartree-Fock(-Bogoliubov) equations~\cite{negele1973neutron, grill2011cluster}.

Ideally, to model the EoS of a NS in a unified way, one should use the same model for all of its layers. The ideal choice is a fully microscopic method based on solving the Hartree-Fock-Bogoliubov~(HFB) equations~\cite{pastore2011superfluid, grill2011cluster, pastore2017new}, using an effective nucleon-nucleon interaction~\cite{skyrme1956cvii, Book:Reinhard2004} adjusted to reproduce selected nuclear observables~\cite{kortelainen2014nuclear}. A big disadvantage of this approach is how the HFB equations treat continuum states in the inner crust using Dirichlet-Neumann boundary conditions~\cite{baldo2006role, margueronEquationStateInner2008, pastore2017new}, which leads to spurious shell effects. New methods to overcome such a problem have been suggested~\cite{chamel2012neutron, jin2017coordinate}, but no systematic calculations of the WS cells have been performed yet.

To avoid this difficulty, several authors have opted to instead use the extended Thomas-Fermi~(ETF) method~\cite{onsiEquationStateStellar1997a,martinLiquidgasCoexistenceEnergy2015,mondalStructureCompositionInner2020}. Due to its semi-classical nature, the ETF method is not affected by the spurious shell effects of the neutron gas encountered in the standard HFB approach.
In Ref.~\cite{onsi2008semi}, the ETF method was extended to use the Strutinsky integral~(SI) correction to recover the important shell effects for the protons in the clusters. This method, named ETFSI, calculates the nuclear energy contribution using parameterised nuclear density profiles, while still using the same energy density functional to generate the fields as in the HFB method. In Ref.~\cite{pearson2015role}, the ETFSI method was further developed to take into account the effects of proton pairing correlations.

In a recent series of articles~\cite{shelley2020accurately, universe6110206}, we have performed a systematic comparison between the HFB and ETFSI methods, which have typically yielded different results for the structure of the inner crust. In particular, we observed that by adding to ETFSI the effects of neutron pairing correlations under a simple local density approximation, the resulting energy per particle agrees well with that obtained using the HFB method, apart from the very low density region where the outer-inner crust transition takes place.

Having quantified the agreement of this ETFSI+pairing method with HFB, we can now address an interesting question, namely, why Zirconium isotopes (i.e. clusters with $Z=40$) are consistently predicted throughout the inner crust~\cite{onsi2008semi, pearsonInnerCrustNeutron2012, pearsonUnifiedEquationsState2018, pearsonErratumUnifiedEquations2019, pearson2020unified, mondalStructureCompositionInner2020}. To answer this, we perform systematic calculations of the inner crust EoS, selecting several Skyrme functionals whose various infinite nuclear matter~(INM) properties~\cite{dutra2012skyrme} cover reasonable ranges. By investigating possible relationships between the INM properties of Skyrme functionals and the proton content and pressure of the WS cells, we aim at providing a better understanding of previous investigations of the structure of the inner crust.

The article is organised as follows: in Sec.~\ref{sec:etfsi}, we outline the main features of the ETFSI+pairing method, and in Sec.~\ref{sec:func} we discuss the Skyrme functionals selected for this work. In Sec.~\ref{sec:inner}, we present results for the chemical composition and EoS of the inner crust, and we give our conclusions in Sec.~\ref{sec:concl}.

%%%%%%%%%%%%%%%%%%%%%%%%%%%%%%%%%%%%%%%%%%%%%%%%%%%%%%%%%%%%%%%%%%%%%%%%%%%%%%%%%%%%%%%%%%%%%%%%%%%%
    
\section{The ETFSI+pairing method}\label{sec:etfsi}

We now briefly review the ETFSI+pairing method, used to study the structure of the inner crust. For a more detailed discussion on the ETFSI method we refer the reader to Refs.~\cite{onsi2008semi,pearsonInnerCrustNeutron2012}.

In this article, we consider only cold non-accreting NS, and so we neglect the effects of temperature~\cite{burrello2015heat, pastore2012superfluid}. Under the WS approximation, for a fixed baryonic density~$n_b$ in the inner crust, we minimise the total energy per particle of the WS cell 

    \begin{equation}\label{eqn:etot}
        e = e_{\mathrm{Sky}} + e_e - Y_p Q_{n,\beta}\;.
    \end{equation}

\noindent $e_{\mathrm{Sky}}$ is the contribution to $e$ from the interaction of the baryons via the strong force and from the Coulomb interaction between the protons. $e_e$ is contribution from the kinetic and potential energies of ultra-relativistic electrons~\cite{shapiro2008black} and from the proton-electron interaction~\cite{grill2011cluster}. The last term accounts for the mass difference between neutrons and protons, $Q_{n,\beta}=0.782$~MeV. $Y_p=Z/A$ is the proton fraction in the cell, where $A$ is the total number of baryons in the cell.

The nuclear part of the energy per particle is expressed as functional of local densities as~\cite{Book:Reinhard2004}

    \begin{eqnarray}\label{eqn:functional}
        e_{\mathrm{Sky}}=\frac{1}{A}\int \mathcal{E}\left(n_q(\mathbf{r}),\tau_q(\mathbf{r}),\vec{J}(\mathbf{r}) \right)d\mathbf{r}\;,
    \end{eqnarray}

\noindent where $q=n,p$ stands for neutrons ($n$) and protons ($p$). Considering only time-reversal invariant systems~\cite{perlinska2004local}, the \emph{standard} Skyrme functional~\cite{becker2017solution} depends only on a linear combination of the matter densities $n_q(r)$, kinetic densities $\tau_q(r)$, and spin current densities $\vec{J}(r)$, and on their derivatives.

To calculate $e_{\mathrm{Sky}}$, we use the ETFSI+pairing method~\cite{onsi2008semi,pearson2015role,universe6110206}. In this approach, the densities $n_q(r)$ are not calculated using the wave-functions, as in the HFB method~\cite{ring2004nuclear}, but are instead parameterised using Fermi-Dirac density profiles~\cite{onsi2008semi} as

    \begin{eqnarray}\label{eqn:fd_profile}
        n_q(r) = \frac{\rho_q^{\mathrm{liq}}-\delta_{q,n}\rho^{\mathrm{gas}}}{1+\exp\left(\frac{r-r_q}{a_q} \right)} + \delta_{q,n}\rho^{\mathrm{gas}}\;.
    \end{eqnarray}

\noindent $\rho_{q=n,p}^{\mathrm{liq}}$ are the densities of the neutrons and protons at the centre of the WS cell, $r=0$, while $\rho^{\mathrm{gas}}$ is the neutron density at the cell edge, $r=R_{\mathrm{WS}}$. $r_{q=n,p}$ are the radii of the neutrons and protons in the cluster, and $a_{q=n,p}$ are their cluster surface diffusivities. These seven adjustable parameters are determined by the minimisation of the energy per particle, given in Eq.~\ref{eqn:etot}, under the constraints of charge neutrality and $\beta$-equilibrium. We also use the relations presented in Ref.~\cite{onsi2008semi}, reducing the number of free parameters to five.
In Ref.~\cite{martinLiquidgasCoexistenceEnergy2015} the authors introduced an exponent for the denominator in Eq.~\ref{eqn:fd_profile}, but they concluded that the results are largely unchanged, so we do not consider such a parameterisation.

The ETF method expresses $\tau_q(r)$ and $\vec{J}(r)$ as a function of the matter densities and their derivatives, up to 4th order in the Wigner-Kirkwood expansion~\cite{Grammaticos1979}. The full expressions of these quantities are given in Ref.~\cite{bartel2002nuclear}.

In this \emph{semiclassical} approach, shell effects are lost; they are recovered with the Strutinsky integral~(SI) correction suggested in Ref.~\cite{onsi2008semi}. The main advantages of this ETFSI method over the HFB method are the reliable treatment of the neutron gas~\cite{baldo2006role,margueronEquationStateInner2008}, and the dramatic reduction in computational cost needed for the energy minimisation. 

In the original ETFSI method~\cite{onsi2008semi,pearsonInnerCrustNeutron2012}, pairing correlations were not taken into account, despite their fundamental role in describing several important phenomena related to the physics of the inner crust~\cite{sandulescu2004nuclear, pastore2011superfluid, watanabe2017superfluid, baldo2005role}. Recent improvements of the model presented in Refs.~\cite{pearson2015role, universe6110206} now allow the inclusion of pairing correlations for both protons and neutrons.

The resulting ETFSI+pairing model leads to results that agree well with the most accurate HFB results currently available~\cite{pastore2017new}. At the interface region between the outer and the inner crust, a non-negligible discrepancy was observed~\cite{universe6110206}, but this concerns a very limited density region and it will not modify the conclusions of this article.

Since Skyrme functionals are often fitted on doubly-magic nuclei, they are not usually equipped with a consistent pairing interaction. Therefore, for each functional used in this work, we add a simple density-dependent pairing interaction of the form~\cite{bertsch1991pair}

    \begin{eqnarray}\label{eqn:pair}
        v^{\mathrm{pair}}_q(\mathbf{r}_1,\mathbf{r}_2)=v_{0q}\left[1-\eta \left(\frac{n_q(r)}{n_0}\right)^\alpha\right]\delta\left(\mathbf{r}_1-\mathbf{r}_2\right)\;.
    \end{eqnarray}

\noindent We choose the parameters $\eta=0.7$ and $\alpha=0.45$. $n_0$ is the saturation density of the functional. We assume that the pairing strength $v_{0q}$ is the same for neutrons and protons, and we fix it to obtain a maximum pairing gap in pure neutron matter~(PNM) of $\approx3$~MeV, as done in Ref.~\cite{grill2011cluster}. These choices may appear arbitrary~\cite{gandolfi2008equation}, but it was shown in Ref.~\cite{universe6110206} that a variation of the pairing strength does not impact the resulting chemical composition of the inner crust. To avoid the ultraviolet divergence of the interaction given in Eq.\ref{eqn:pair}~\cite{bulgac2002renormalization}, we adopt a smooth cut-off in quasi-particle space as defined in Ref.~\cite{grill2011cluster}.

For functionals of the BSk family~\cite{goriely2013further}, we keep the pairing interaction used and developed by the authors. 

%%%%%%%%%%%%%%%%%%%%%%%%%%%%%%%%%%%%%%%%%%%%%%%%%%%%%%%%%%%%%%%%%%%%%%%%%%%%%%%%%%%%%%%%%%%%%%%%%%%%

\section{Choice of functionals}\label{sec:func}

According to Ref.~\cite{dutra2012skyrme}, more than 240 Skyrme functionals have been published in the literature. Typically, their parameters are adjusted to reproduce the binding energy of some (or all) atomic nuclei, and also some INM properties. There is no standard parameter fitting protocol; various (pseudo)-observables are used in fitting, with widely varying uncertainties. As a consequence, it is difficult to assess the quality of functionals.

    \begin{figure}
        \centering
        \includegraphics{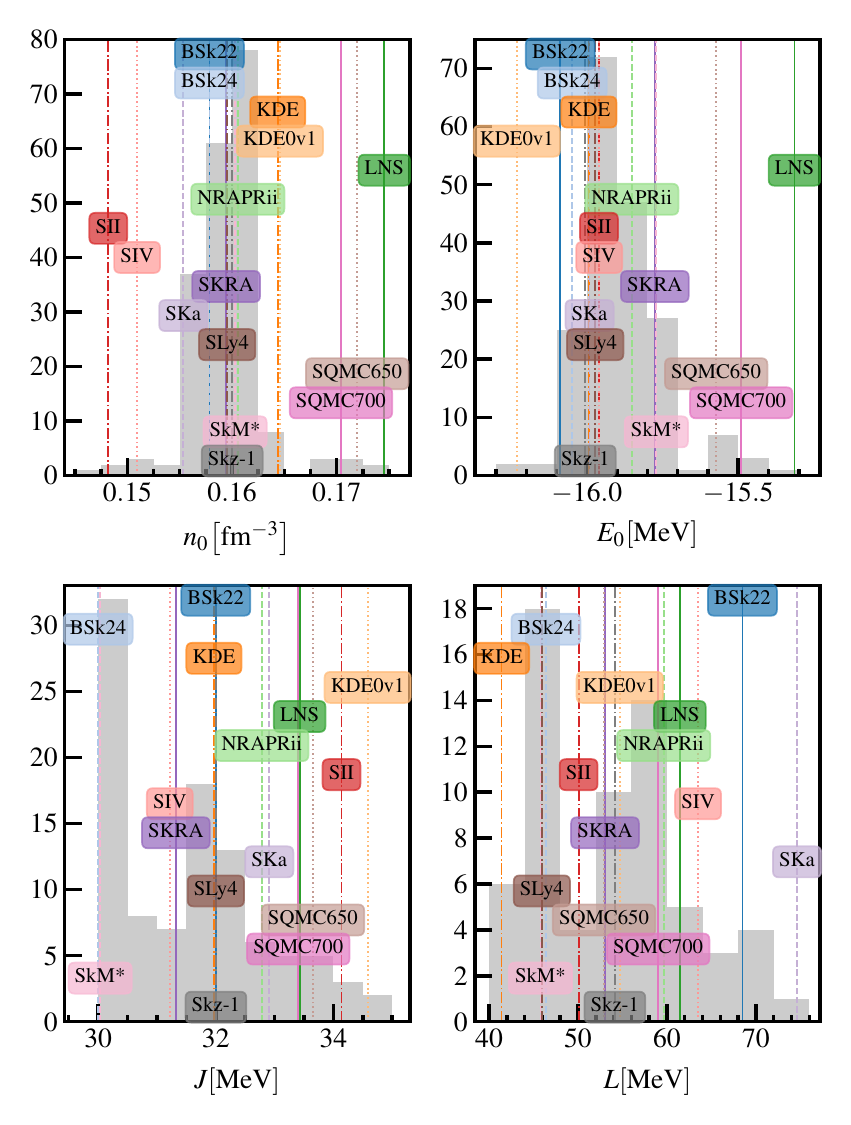}
        \caption{(Colors online) Each panel represents the distribution of the Skyrme functionals given in Ref.~\cite{dutra2012skyrme} for four INM properties. Starting from the top left panel and going clockwise we show $n_0$, $E_0$, $L$ and $J$. The values of each quantity for the 15 functionals selected in this work are shown as labelled vertical lines. See text for details.}
        \label{fig:histo}
    \end{figure}

The neutron gas in the WS cell is similar to an infinite nuclear system, and comprises the majority of the baryonic matter in the inner crust. We therefore focus on four INM properties of the functionals: the saturation density $n_0$, the energy per particle $E_0$, the symmetry energy $J$, and its slope $L$, all evaluated at $n_0$. These quantities depend exactly on the EoS of INM~\cite{piekarewicz2009incompressibility}, and so one can expect that variations in the EoS lead to modifications in the energy per particle of the neutron gas, thus changing the relative energy contributions from baryons and electrons in the WS cell, and ultimately affecting the chemical composition and pressure in the inner crust. In Fig.~\ref{fig:histo}, we show as histograms the distributions of $n_0$, $E_0$, $J$, and $L$, for all the functionals in Ref.~\cite{dutra2012skyrme}.

By inspecting the lower panels of Fig.~\ref{fig:histo}, we observe that not all functionals given in Ref.~\cite{dutra2012skyrme} are shown. In our selection process, we imposed two additional constraints on the symmetry energy and its slope, namely $30\leq J\leq35$~MeV and $L=58\pm18$~MeV. These ranges of values are the ones suggested in Ref.~\cite{dutra2012skyrme}. These constraints are to some extent arbitrary, as there is no consensus on what constitutes a reasonable range of values for $J$ and $L$, but they are consistent with the findings of recent analyses~\cite{tsang2012constraints, chen2010density, vidana2009density, roca2011neutron, liTopicalIssueNuclear2014}, and with results and associated uncertainties from new chiral~EFT calculations~\cite{drischlerChiralInteractionsNexttoNexttoNexttoLeading2019,drischlerHowWellWe2020}.

From this group, we finally selected 15 functionals which have a wide range of values for each of $n_0$, $E_0$, $J$, and $L$: BSk22 and BSk24~\cite{goriely2013further}, KDE~\cite{agrawal2005determination}, KDE0v1~\cite{agrawal2005determination}, LNS~\cite{cao2006brueckner}, NRAPRii~\cite{stevenson2013skyrme} (NRAPR~\cite{steiner2005isospin} with the spin-orbit strength modified), SII~\cite{vautherin1972hartree}, SIV~\cite{beiner1975nuclear}, SKRA~\cite{rashdan2000skyrme}, SKA~\cite{kohler1976skyrme}, SLy4~\cite{chabanat1998skyrme}, SQMC650 and SQMC700~\cite{guichon2006physical}, SkM$^*$~\cite{krivine1980derivation}, and Skz-1~\cite{margueron2002instabilities}. Their INM properties are labelled in Fig.~\ref{fig:histo}. Five of them --- \textbf{KDE0v1}, \textbf{LNS}, \textbf{NRAPR}, \textbf{SKRA} and \textbf{SQMC700} --- are consistent with all the INM constraints presented in Ref.~\cite{dutra2012skyrme}, and are named in that reference and hereafter as the CSkP$^*$ set.

    \begin{figure}
        \centering
        \includegraphics{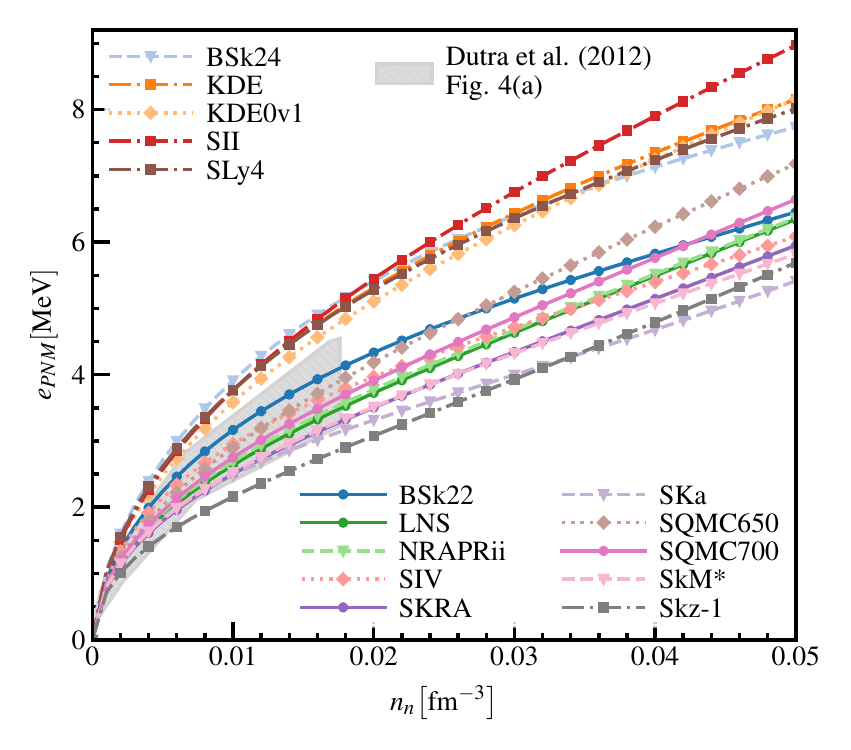}
        \caption{(Colors online) The EoS of PNM as a function of the neutron density $n_n$ for the 15 functionals selected in this work. The shaded area represents the constraint given in Fig.~4(a) of Ref.~\cite{dutra2012skyrme}. The functionals labelled in the upper left are the Stiff set, and those in the lower right the Soft set. See text for details.}
        \label{fig:ePNM}
    \end{figure}
   
For our functional selection, we also show the energy per particle~$e_{\mathrm{PNM}}$ in PNM in Fig.~\ref{fig:ePNM}, for the density range relevant for spherical inner crust calculations. The functionals clearly fall into two distinct families: the set BSk24, KDE, KDE0v1, SII, and SLy4, with a stiff EoS at these low densities (hereafter the \emph{Stiff} set), and the set BSk22, LNS, NRAPRii, SIV, SKRA, SKa, SQMC650, SQMC700, SkM*, and Skz-1, with a very soft EoS (hereafter the \emph{Soft} set). On the same figure, we have added a shaded area which corresponds to the range spanned by several \emph{ab-initio} calculations used to derive the EoS in PNM, and which is discussed in Fig.~4(a) of Ref.~\cite{dutra2012skyrme}. 

The Stiff set, including SLy4 and BSk24 which have been widely used for NS calculations, are in disagreement with the \emph{ab-initio} calculations at these densities. In contrast, the majority of the Soft set are in reasonable agreement, apart from Skz-1 which is very soft. However, the \emph{ab-initio} calculations selected in Ref.~\cite{dutra2012skyrme} do not represent the entire range available in the literature. As shown in Ref.\cite{goriely2013further}, BSk24 and SLy4 are compatible with the error bars provided by chiral effective field theory calculations from Ref.~\cite{Tews}.

    \begin{figure}
        \centering
        \includegraphics{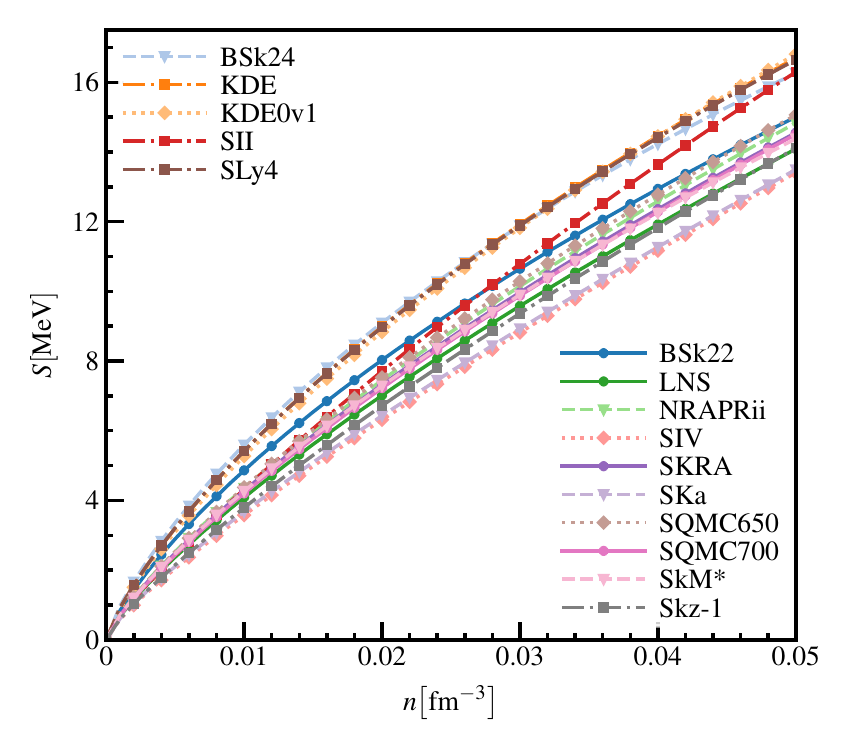}
        \caption{(Colors online) The symmetry energy $S$ as a function of density for selected functionals. The functionals labelled in the upper left are the Stiff set, and those in the lower right the Soft set. See text for details.}
        \label{fig:Esymn}
    \end{figure}

The inner crust does not only comprise neutrons, but also has a significant proton fraction, and so it is interesting to observe the behaviour of the symmetry energy~$S$ in the relevant density region and not only at $n_0$. This is shown in Fig.~\ref{fig:Esymn}, where we again observe a similar grouping into the Stiff and Soft sets as shown in Fig.~\ref{fig:ePNM}. The reason is quite simple and can be understood from Fig.~\ref{fig:histo}: the vast majority of the functionals have values of $n_0$ and $E_0$ in a very narrow range. This means that all these functionals have a very similar EoS in symmetric nuclear matter (SNM) in the low density region. Since $S$ is just the difference between the EoS in PNM and SNM (within the parabolic approximation~\cite{davesne2016extended}), it follows that the pattern observed in PNM repeats here in a very similar way. The exception to this simple rule are SII and SIV. They have extremely low values of $n_0$, shown in Fig.~\ref{fig:histo}, and so a significantly different EoS in SNM compared to the other functionals.

%%%%%%%%%%%%%%%%%%%%%%%%%%%%%%%%%%%%%%%%%%%%%%%%%%%%%%%%%%%%%%%%%%%%%%%%%%%%%%%%%%%%%%%%%%%%%%%%%%%%

\section{Inner crust composition}\label{sec:inner}
To calculate the EoS and determine the chemical composition of the inner crust, we minimise the total energy per particle~$e$ in the WS cell, given in Eq.\ref{eqn:etot}, using the method explained in Sec.~\ref{sec:etfsi} and Ref.~\cite{universe6110206}. We cover the range of baryonic densities ${n_b\in[0.00025,0.05]}$~fm$^{-3}$, above which non-spherical pasta phases are expected to appear~\cite{pearson2020unified}, and the range of proton numbers ${Z\in[16,60]}$.

    \begin{figure*}
        \centering
        \includegraphics{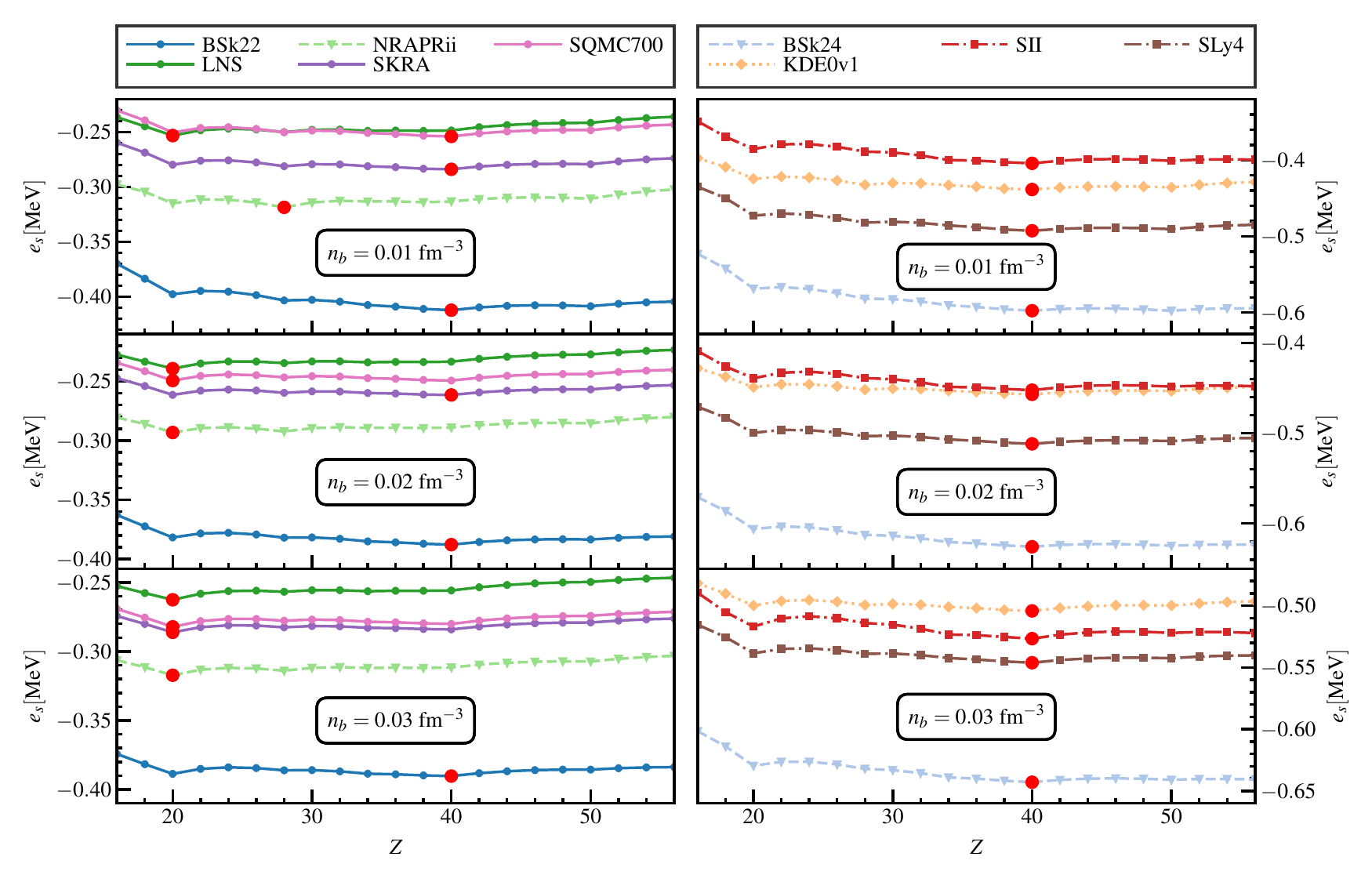}
        \caption{(Colors online) The re-scaled energy per particle $e_s$, defined in Eq.\ref{escale}, for selected functionals from the Soft (left column) and Stiff (right column) sets, at three values of baryonic density~$n_b$. The large dot on each curve represents the position of the energy minimum. See text for details.}
        \label{fig:slices}
    \end{figure*}

We first show results for a few selected baryonic densities $n_b=0.01, 0.02$ and $0.03$~fm$^{-3}$, for a smaller group of functionals. Figure~\ref{fig:slices} illustrates the energy per particle as a function of $Z$. We show a re-scaled energy per particle, $e_s$, defined as

    \begin{eqnarray}\label{escale}
        e_s(n_b) = e(n_b) - e_{\mathrm{PNM}}(n_b) - e_{n,\mathrm{pair}}(n_b)\;,
    \end{eqnarray}

\noindent where $e_{n,\mathrm{pair}}$ is the neutron pairing energy per particle. This is purely for visual reasons, so that all values lie within a similar energy range. Since $e_{\mathrm{PNM}}$ is independent of $Z$, and $e_{n,\mathrm{pair}}$ is roughly constant with respect to $Z$~\cite{universe6110206}, this results in a simple shift in $e$ for a given functional. In the left column are five from the Soft set: \textbf{BSk22}, \textbf{LNS}, \textbf{NRAPRii}, \textbf{SKRA}, and \textbf{SQMC700}; in the right column are four functionals from the Stiff set: \textbf{BSk24}, \textbf{KDE0v1}, \textbf{SII}, and \textbf{SLy4}, as explained in Sec.~\ref{sec:func}.

In this figure, we clearly see the different behaviour of the Soft and Stiff sets. The Stiff set has a persistent minimum at $Z=40$, while the Soft set has minima that favour lower $Z$, or that shift towards lower $Z$ as $n_b$ increases, with all favouring $Z=20$ at $n_b=0.03$~fm$^{-3}$. The exception in the Soft set is BSk22, which transitions to $Z=20$ just above $n_b=0.03$~fm$^{-3}$, and which has one of the stiffest PNM EoS in this set. NRAPRii is unique among all the functionals investigated, in that it favours $Z=28$ at $n_b=0.01$~fm$^{-3}$. This is likely related to an issue involving the spin-orbit parameter, $W_0$. See the discussion in Ref.~\cite{stevenson2013skyrme} for more details.
Below $n_b=0.01$~fm$^{-3}$, nearer the transition region between the outer and inner crust, other finite-size effects not considered in our ETFSI+pairing method may become significant, changing the results. The HFB method is preferable in this very low density range.

    \begin{figure}
        \centering
        \includegraphics{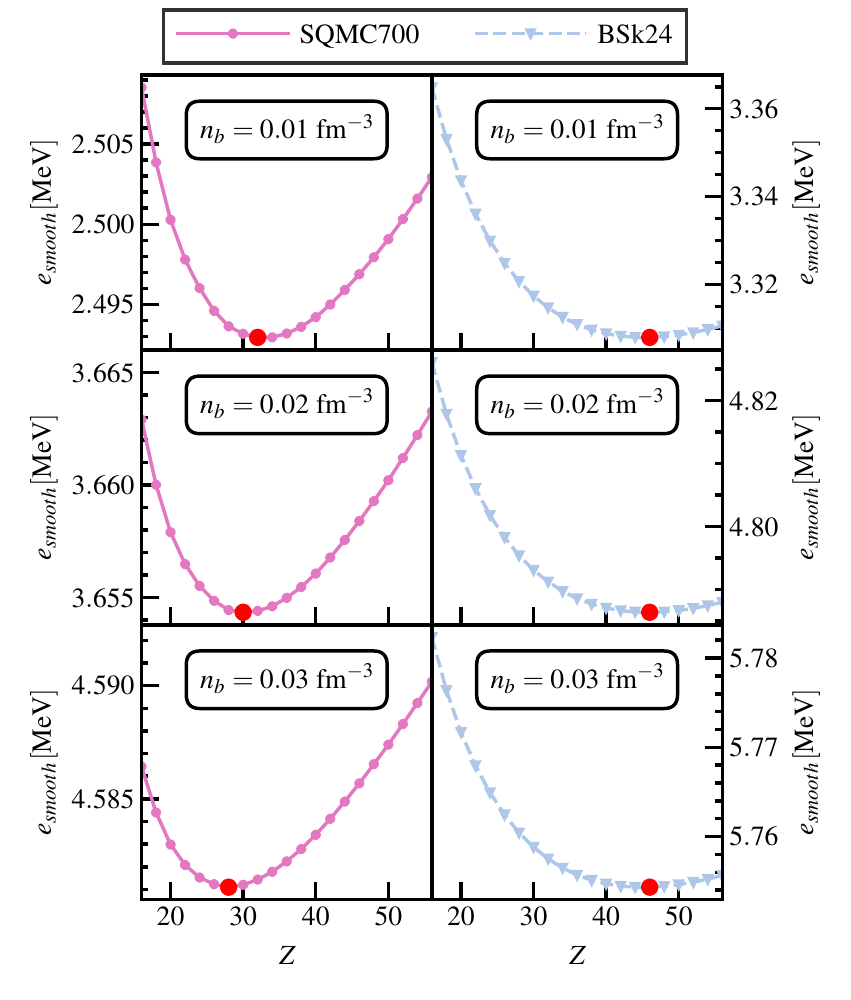}
        \caption{(Colors online) Total energy per particle, $e_{\mathrm{smooth}}$, obtained without using shell or pairing corrections, for two functionals at three representative baryonic densities~$n_b$. See text for details.}
        \label{fig:smoothSlices}
    \end{figure}
    
To better understand the origin of the different minima, we plot in Fig.~\ref{fig:smoothSlices} the \emph{smooth} part of the total energy per particle, $e_{\mathrm{smooth}}$, defined as

    \begin{equation}
        e_{\mathrm{smooth}} = e - e_{n,\mathrm{pair}} - (e_p^{SI}+e_{p,\mathrm{pair}})\;,
    \end{equation}

\noindent where the last term is the sum of the SI correction with BCS pairing for protons, and the proton pairing energy, as explained in Ref.~\cite{pearson2015role}. This results in a smooth parabola-like shape, with a single minimum. We select two representative functionals \textbf{SQMC700} and \textbf{BSk24}, from the Soft and Stiff sets respectively. The parabola shifts significantly in $Z$ going between the two functionals. Furthermore, as $n_b$ increases, we observe that for SQMC700 the minimum moves from $Z=32$ to $Z=28$, while for BSk24 it stays at $Z=46$. 

When the SI correction is included, as in Fig.~\ref{fig:slices}, each functional at each density displays local minima at (semi\nobreakdash-)magic $Z$ values between $20$ and $50$. However, Fig.~\ref{fig:smoothSlices} shows how the global minimum in each case is governed more by the stiffness of the PNM EoS.
In this work, we strictly consider no temperature effects. Since the various energy minima are quite close to each other, the inclusion of such effects may change this picture. See discussion in Ref.~\cite{burrello2015heat} for more details.

    \begin{figure}
        \centering
        \includegraphics{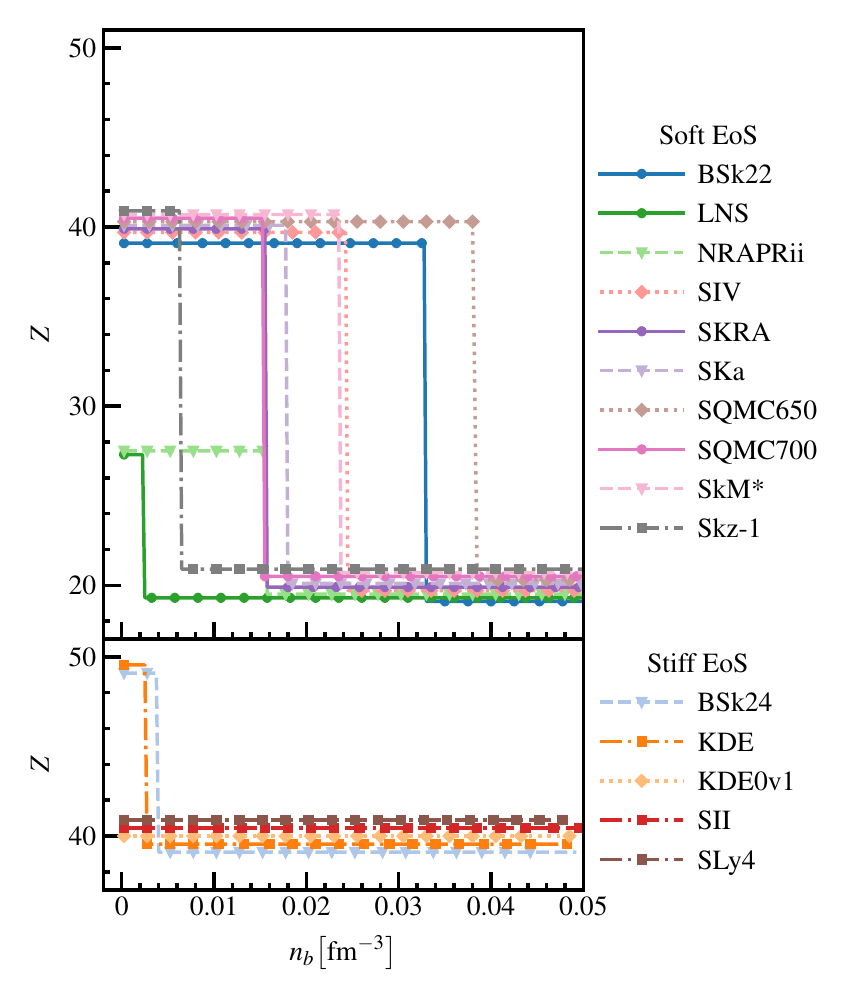}
        \caption{(Colors online) Proton content of WS cells in the inner crust as a function of the baryonic density $n_b$, for all functionals in the Soft and Stiff sets. A small vertical offset has been added to each line to make them more visible; all lines only take values from one of $Z=20, 28, 40, 50$. See text for details.}
        \label{fig:EoS_Z}
    \end{figure}
  
A more systematic study of the evolution of the proton content of the clusters in the inner crust is illustrated in Fig.~\ref{fig:EoS_Z}, for all functionals in the Soft and Stiff sets. For clarity, small vertical offsets have been added to each line, and the Soft and Stiff sets in are shown in separate panels. At very low $n_b$ several values of Z are observed, but above $n_b=0.02$~fm$^{-3}$ only two minima are observed: $Z=20$ and $Z=40$. The minimum at $Z=40$ seems to be favoured by most functionals at around $n_b=0.01$~fm$^{-3}$, but at $n_b=0.03$~fm$^{-3}$ the majority of the Soft set have transitioned to the $Z=20$ configuration. The only exception to this rough classification is SQMC650, whose $e_{\mathrm{PNM}}$ is the highest out of the Soft set, as seen in Fig.~\ref{fig:ePNM}, and it maintains a $Z=40$ minimum up to quite a high baryonic density of $n_b\approx0.04$~fm$^{-3}$.

It is interesting to note that $Z=20$ is favoured by functionals whose PNM EoS is compatible with \emph{ab-initio} calculations. This result is also in good agreement with the findings of Ref.~\cite{baldo2005role}. Although further analysis is necessary, our results suggest that a better understanding of the EoS in low-density PNM may help to clarify the chemical composition of the inner crust.

    \begin{figure}
        \centering
        \includegraphics{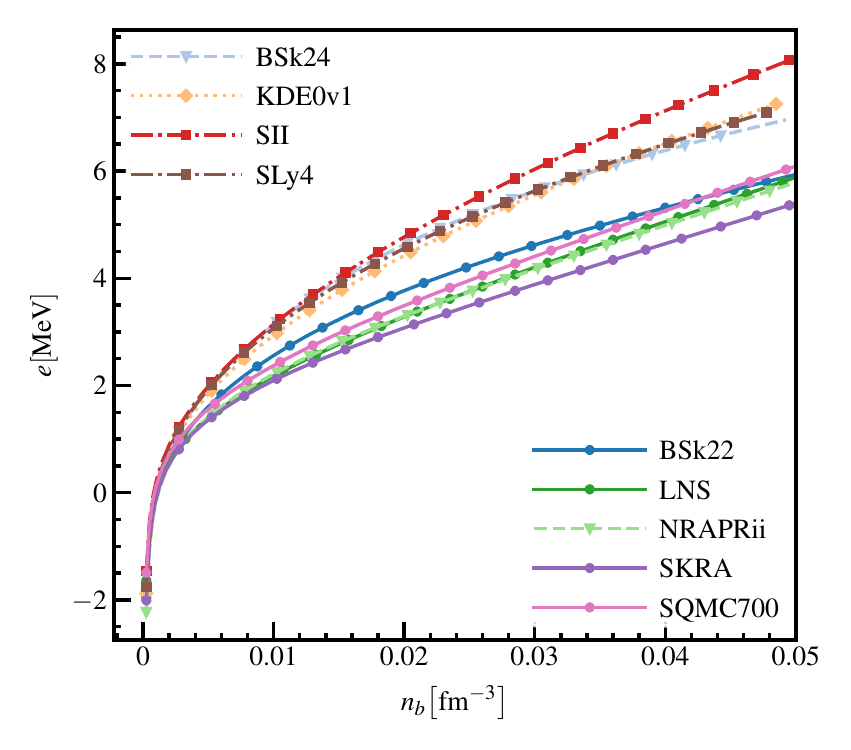}
        \caption{(Colors online) The energy per particle $e$ in the inner crust, as a function of the baryonic density $n_b$. Four functionals from the Stiff set are shown, labelled in the upper left, and five from the Soft set, labelled in the lower right. See text for details. }
        \label{fig:e_crust}
    \end{figure}

Having calculated the chemical composition of the inner crust, we now study its general properties with the different functionals. In Fig.~\ref{fig:e_crust}, we show the total energy per particle~$e$ (Eq.~\ref{eqn:etot}) for the WS cells in the inner crust, obtained through the energy minimisation, as a function of the baryonic density~$n_b$. The pattern formed by the various Soft and Stiff functionals is almost identical to that seen for $e_{\mathrm{PNM}}$, shown in Fig.~\ref{fig:ePNM}. This further supports the PNM EoS being the major driver of the inner crust EoS.

    \begin{figure}
        \centering
        \includegraphics{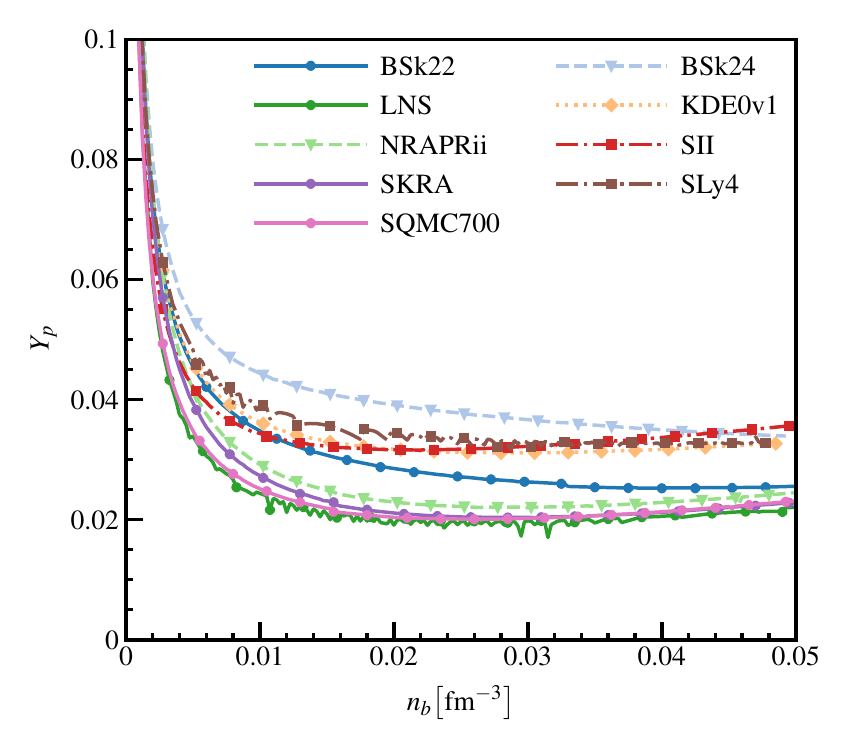}
        \caption{(Colors online) Proton fraction $Y_p$ in the inner crust, as a function of the baryonic density $n_b$. Four functionals from the Stiff set are shown, labelled in the upper right, and five from the Soft set, labelled in the upper left. See text for details.}
        \label{fig:Yp}
    \end{figure}

Another relevant astrophysical quantity is the proton fraction in the inner crust, $Y_p=Z/A$. It plays a role in the transport properties of a NS and in the determination of the neutrino mean free path in the stellar medium~\cite{iwamoto1982effects}. In Fig.~\ref{fig:Yp}, we report $Y_p$ as a function of $n_b$ for the same smaller selection of functionals shown in Fig.~\ref{fig:slices}. The same feature already observed in Fig.~\ref{fig:EoS_Z} is clear: the Soft functionals predict WS cells with fewer protons, and have a lower $Y_p$, while the contrary is true for the Stiff functionals. At around $n_b=0.02$~fm$^{-3}$, $Y_p$ varies by up to a factor of 2 between the two sets. When $Z$ in a WS cell changes, the baryon content of each WS cell follows this change quite closely. As a consequence, the factor of 2 observed between the two dominant minima in Fig.~\ref{fig:EoS_Z} ($Z=20,40$) is conserved here.

It is worth commenting on the case of BSk24, a functional specifically created to be able to provide a unified description of the NS EoS. This was adjusted under many constraints, including the requirement that it reproduce the LS2 equation of state~\cite{liNeutronStarStructure2008} in PNM, which was calculated using the microscopic Brueckner-Hartree-Fock method. The authors note in Ref.~\cite{goriely2013further} that they focused on supernuclear densities when constraining BSk24 to the LS2 EoS. As a result, the BSk24 PNM EoS is significantly stiffer than LS2 at inner crust densities. Around $n_b=0.04$~fm$^{-3}$, LS2 is better approximated by SQMC650, SQMC700, and NRAPRii, than by BSk24.

In Refs.~\cite{pearsonUnifiedEquationsState2018,pearsonErratumUnifiedEquations2019}, the authors claim that the constraining PNM EoS has little effect on $Y_p$ in the inner crust, and show that $J$ is an important INM quantity. However, by inspecting carefully Figs.~\ref{fig:ePNM} and~\ref{fig:Yp}, we see that BSk24 ($J=30$~MeV), SLy4 ($J=32$~MeV), and KDE0v1 ($J=34.6$~MeV) have a $Y_p$ largely in the range $0.03-0.04$. The functionals with a softer PNM EoS --- BSk22, LNS, NRAPRii, SKRA, and SQMC700 --- have a $Y_p$ mostly in the range $0.02-0.025$, despite having $J$ ranging from $31.3$ to $33.4$~MeV. We therefore conclude that $J$ is not the driving factor leading to large variations of $Y_p$, but it is the equation of state of PNM at low density, or equivalently the symmetry energy $S$ at low density as shown in Fig.~\ref{fig:Esymn} and discussed in Sec.~\ref{sec:func}.

The softest functional investigated in Ref.~\cite{pearsonUnifiedEquationsState2018, pearsonErratumUnifiedEquations2019}, BSk22, which more closely follows LS2 at inner crust densities, gives results favouring an energy minimum at $Z=20$ at intermediate inner crust densities, and lower values for $Y_p$, $e$, and $P$. This means that the results presented in this article are consistent with those in Refs.~\cite{pearsonUnifiedEquationsState2018,pearsonErratumUnifiedEquations2019}, but given the small variations in the properties of the functionals used, the authors were not able to observe such an interesting correlation. 

It is also  interesting to note from Fig.~\ref{fig:Yp} that almost all of the CSkP$^*$ set, selected in Ref.~\cite{dutra2012skyrme} according to several criteria based on INM properties, tend to favour a very small $Y_p$ within the crust. The only exception is KDE0v1, from the Stiff set. 

    \begin{figure}
        \centering
        \includegraphics{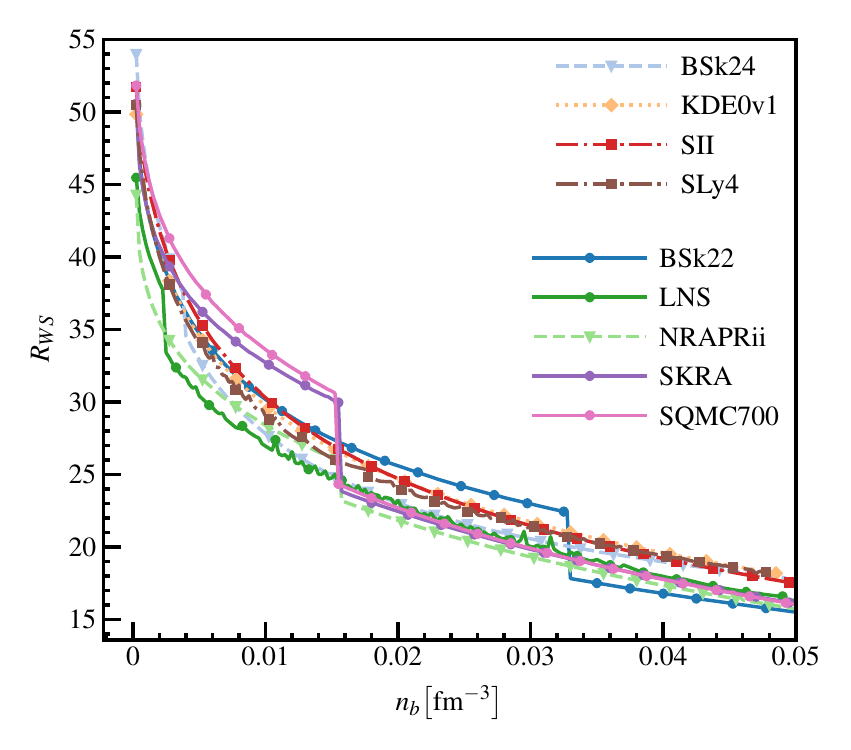}
        \caption{(Colors online) Radius of WS cells $R_{\mathrm{WS}}$ in the inner crust, as a function of the baryonic density $n_b$. Four functionals from the Stiff set are shown, labelled in the upper right, and five from the Soft set, labelled just below on the right. See text for details.}
        \label{fig:RWS}
    \end{figure}

In Fig.~\ref{fig:RWS}, we illustrate the evolution of the radius of the Wigner-Seitz cell, $R_{\mathrm{WS}}$, as a function of the baryonic density, again for the smaller selection of functionals.
While in the low density region, ${n_b\in[0.00025,0.015]}$~fm$^{-3}$, we observe significant variations in the sizes of the cells, at higher baryonic densities, we notice that the size of the cell is almost independent of the functional. In particular, we clearly identify again the Stiff set and the Soft set, giving values of $R_{\mathrm{WS}}$ very close to each other.
The sudden jumps shown in this figure correspond to abrupt changes in the chemical composition of the crust, as seen in Fig.~\ref{fig:EoS_Z}.

    \begin{figure}
        \centering
        \includegraphics{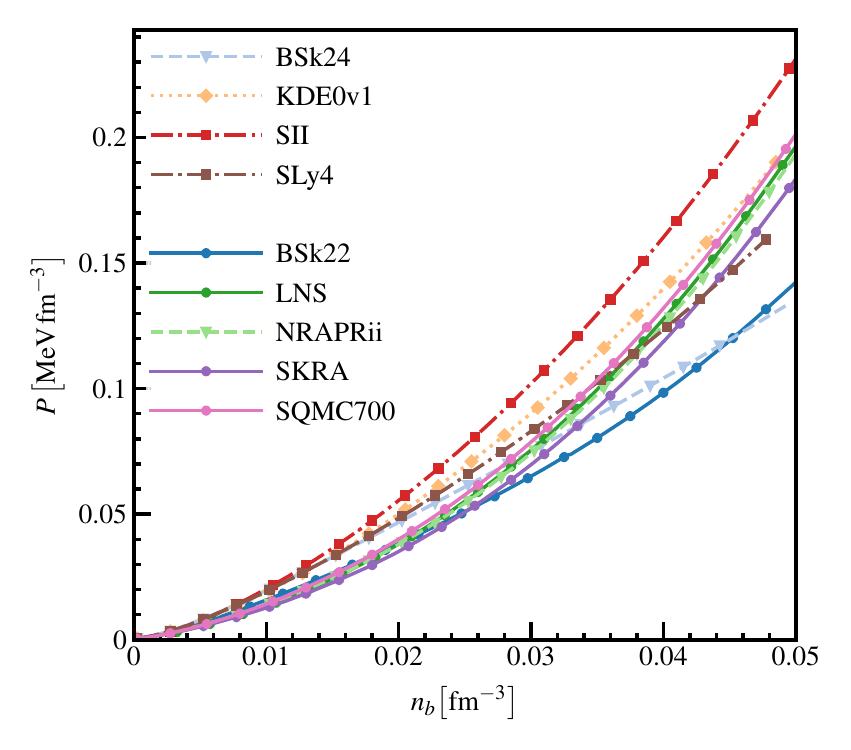}
        \caption{Pressure $P$ in the inner crust, as a function of the baryonic density $n_b$. Four functionals from the Stiff set are shown, labelled in the upper left, and five from the Soft set, labelled just below on the left. See text for details.}
        \label{fig:pressure}
    \end{figure}

Finally, in Fig.~\ref{fig:pressure}, we illustrate the evolution of the pressure~$P$ as a function of the baryonic density~$n_b$, for the same eight functionals already discussed in Figs.~\ref{fig:e_crust} and~\ref{fig:Yp}. In this case, the separation into the Stiff and Soft sets is only visible up to around $n_b=0.025$~fm$^{-3}$. We recall that the pressure depends on the derivatives of the EoS, rather than its value.

%%%%%%%%%%%%%%%%%%%%%%%%%%%%%%%%%%%%%%%%%%%%%%%%%%%%%%%%%%%%%%%%%%%%%%%%%%%%%%%%%%%%%%%%%%%%%%%%%%%%

\section{Conclusions}\label{sec:concl}

In this article we have performed a systematic analysis of the structure of the inner crust of a non-accreting NS, using the ETFSI+pairing method.
By selecting 15 different functionals with different infinite nuclear matter properties, we have illustrated a possible correlation between the a functional's PNM EoS at low density, and the proton content and pressure of the Wigner-Seitz cells.
In particular, we have shown that functionals with a \emph{soft} PNM EoS at low densities tend to favour the $Z=20$ minimum, as well as a lower proton fraction, and a lower pressure up to around $n_b=0.025$~fm$^{-3}$, while functionals with a \emph{stiff} PNM EoS show the opposite behaviour, and favour instead the $Z=40$ minimum.
This result clearly shows the importance of the constraining the PNM EoS more tightly at subnuclear densities in the adjustment protocol of Skyrme functionals used to study NS.
This is not the \emph{standard} procedure, since authors tend to focus more on the high density trend of the EoS; indeed, this is necessary to prevent the collapse of the NS, and to reproduce its global properties such as the maximum mass and radius~\cite{zdunik2017neutron}.
More stringent constraints on functionals obtained by more recent \emph{ab-initio} calculations at low density should pave the way toward a more reliable unified EoS for neutron star matter.

%%%%%%%%%%%%%%%%%%%%%%%%%%%%%%%%%%%%%%%%%%%%%%%%%%%%%%%%%%%%%%%%%%%%%%%%%%%%%%%%%%%%%%%%%%%%%%%%%%%%

\section*{Acknowledgments}

We thank M. Urban for the interesting discussions that motivated this article. This work was supported by STFC Grant No. ST/P003885/1.

%%%%%%%%%%%%%%%%%%%%%%%%%%%%%%%%%%%%%%%%%%%%%%%%%%%%%%%%%%%%%%%%%%%%%%%%%%%%%%%%%%%%%%%%%%%%%%%%%%%%

\bibliography{bib}

\begin{thebibliography}{86}%
\makeatletter
\providecommand \@ifxundefined [1]{%
 \@ifx{#1\undefined}
}%
\providecommand \@ifnum [1]{%
 \ifnum #1\expandafter \@firstoftwo
 \else \expandafter \@secondoftwo
 \fi
}%
\providecommand \@ifx [1]{%
 \ifx #1\expandafter \@firstoftwo
 \else \expandafter \@secondoftwo
 \fi
}%
\providecommand \natexlab [1]{#1}%
\providecommand \enquote  [1]{``#1''}%
\providecommand \bibnamefont  [1]{#1}%
\providecommand \bibfnamefont [1]{#1}%
\providecommand \citenamefont [1]{#1}%
\providecommand \href@noop [0]{\@secondoftwo}%
\providecommand \href [0]{\begingroup \@sanitize@url \@href}%
\providecommand \@href[1]{\@@startlink{#1}\@@href}%
\providecommand \@@href[1]{\endgroup#1\@@endlink}%
\providecommand \@sanitize@url [0]{\catcode `\\12\catcode `\$12\catcode
  `\&12\catcode `\#12\catcode `\^12\catcode `\_12\catcode `\%12\relax}%
\providecommand \@@startlink[1]{}%
\providecommand \@@endlink[0]{}%
\providecommand \url  [0]{\begingroup\@sanitize@url \@url }%
\providecommand \@url [1]{\endgroup\@href {#1}{\urlprefix }}%
\providecommand \urlprefix  [0]{URL }%
\providecommand \Eprint [0]{\href }%
\providecommand \doibase [0]{https://doi.org/}%
\providecommand \selectlanguage [0]{\@gobble}%
\providecommand \bibinfo  [0]{\@secondoftwo}%
\providecommand \bibfield  [0]{\@secondoftwo}%
\providecommand \translation [1]{[#1]}%
\providecommand \BibitemOpen [0]{}%
\providecommand \bibitemStop [0]{}%
\providecommand \bibitemNoStop [0]{.\EOS\space}%
\providecommand \EOS [0]{\spacefactor3000\relax}%
\providecommand \BibitemShut  [1]{\csname bibitem#1\endcsname}%
\let\auto@bib@innerbib\@empty
%</preamble>
\bibitem [{\citenamefont {Chamel}\ and\ \citenamefont
  {Haensel}(2008)}]{chamel2008physics}%
  \BibitemOpen
  \bibfield  {author} {\bibinfo {author} {\bibfnamefont {N.}~\bibnamefont
  {Chamel}}\ and\ \bibinfo {author} {\bibfnamefont {P.}~\bibnamefont
  {Haensel}},\ }\bibfield  {title} {\bibinfo {title} {Physics of neutron star
  crusts},\ }\href@noop {} {\bibfield  {journal} {\bibinfo  {journal} {Living
  Reviews in relativity}\ }\textbf {\bibinfo {volume} {11}},\ \bibinfo {pages}
  {10} (\bibinfo {year} {2008})}\BibitemShut {NoStop}%
\bibitem [{\citenamefont {Mantziris}\ \emph {et~al.}(2020)\citenamefont
  {Mantziris}, \citenamefont {Pastore}, \citenamefont {Vidana}, \citenamefont
  {Watts}, \citenamefont {Bashkanov},\ and\ \citenamefont
  {Romero}}]{mantziris2020neutron}%
  \BibitemOpen
  \bibfield  {author} {\bibinfo {author} {\bibfnamefont {A.}~\bibnamefont
  {Mantziris}}, \bibinfo {author} {\bibfnamefont {A.}~\bibnamefont {Pastore}},
  \bibinfo {author} {\bibfnamefont {I.}~\bibnamefont {Vidana}}, \bibinfo
  {author} {\bibfnamefont {D.}~\bibnamefont {Watts}}, \bibinfo {author}
  {\bibfnamefont {M.}~\bibnamefont {Bashkanov}},\ and\ \bibinfo {author}
  {\bibfnamefont {A.}~\bibnamefont {Romero}},\ }\bibfield  {title} {\bibinfo
  {title} {Neutron star matter equation of state including $ d^{*} $-hexaquark
  degrees of freedom},\ }\href@noop {} {\bibfield  {journal} {\bibinfo
  {journal} {Astronomy \& Astrophysics}\ }\textbf {\bibinfo {volume} {638}},\
  \bibinfo {pages} {A40} (\bibinfo {year} {2020})}\BibitemShut {NoStop}%
\bibitem [{\citenamefont {Xu}\ \emph {et~al.}(2009)\citenamefont {Xu},
  \citenamefont {Chen}, \citenamefont {Li},\ and\ \citenamefont
  {Ma}}]{xu2009locating}%
  \BibitemOpen
  \bibfield  {author} {\bibinfo {author} {\bibfnamefont {J.}~\bibnamefont
  {Xu}}, \bibinfo {author} {\bibfnamefont {L.-W.}\ \bibnamefont {Chen}},
  \bibinfo {author} {\bibfnamefont {B.-A.}\ \bibnamefont {Li}},\ and\ \bibinfo
  {author} {\bibfnamefont {H.-R.}\ \bibnamefont {Ma}},\ }\bibfield  {title}
  {\bibinfo {title} {Locating the inner edge of the neutron star crust using
  terrestrial nuclear laboratory data},\ }\href@noop {} {\bibfield  {journal}
  {\bibinfo  {journal} {Physical Review C}\ }\textbf {\bibinfo {volume} {79}},\
  \bibinfo {pages} {035802} (\bibinfo {year} {2009})}\BibitemShut {NoStop}%
\bibitem [{\citenamefont {Lattimer}\ \emph {et~al.}(1977)\citenamefont
  {Lattimer}, \citenamefont {Mackie}, \citenamefont {Ravenhall},\ and\
  \citenamefont {Schramm}}]{lattimerDecompressionColdNeutron1977}%
  \BibitemOpen
  \bibfield  {author} {\bibinfo {author} {\bibfnamefont {J.~M.}\ \bibnamefont
  {Lattimer}}, \bibinfo {author} {\bibfnamefont {F.}~\bibnamefont {Mackie}},
  \bibinfo {author} {\bibfnamefont {D.~G.}\ \bibnamefont {Ravenhall}},\ and\
  \bibinfo {author} {\bibfnamefont {D.~N.}\ \bibnamefont {Schramm}},\
  }\bibfield  {title} {\bibinfo {title} {The decompression of cold neutron star
  matter},\ }\href {https://doi.org/10.1086/155148} {\bibfield  {journal}
  {\bibinfo  {journal} {The Astrophysical Journal}\ }\textbf {\bibinfo {volume}
  {213}},\ \bibinfo {pages} {225} (\bibinfo {year} {1977})}\BibitemShut
  {NoStop}%
\bibitem [{\citenamefont {Tsang}\ \emph
  {et~al.}(2012{\natexlab{a}})\citenamefont {Tsang}, \citenamefont {Read},
  \citenamefont {Hinderer}, \citenamefont {Piro},\ and\ \citenamefont
  {Bondarescu}}]{tsangResonantShatteringNeutron2012}%
  \BibitemOpen
  \bibfield  {author} {\bibinfo {author} {\bibfnamefont {D.}~\bibnamefont
  {Tsang}}, \bibinfo {author} {\bibfnamefont {J.~S.}\ \bibnamefont {Read}},
  \bibinfo {author} {\bibfnamefont {T.}~\bibnamefont {Hinderer}}, \bibinfo
  {author} {\bibfnamefont {A.~L.}\ \bibnamefont {Piro}},\ and\ \bibinfo
  {author} {\bibfnamefont {R.}~\bibnamefont {Bondarescu}},\ }\bibfield  {title}
  {\bibinfo {title} {Resonant {{Shattering}} of {{Neutron Star Crusts}}},\
  }\href {https://doi.org/10.1103/PhysRevLett.108.011102} {\bibfield  {journal}
  {\bibinfo  {journal} {Physical Review Letters}\ }\textbf {\bibinfo {volume}
  {108}},\ \bibinfo {pages} {011102} (\bibinfo {year}
  {2012}{\natexlab{a}})}\BibitemShut {NoStop}%
\bibitem [{\citenamefont {Thompson}\ and\ \citenamefont
  {Duncan}(1995)}]{thompson1995soft}%
  \BibitemOpen
  \bibfield  {author} {\bibinfo {author} {\bibfnamefont {C.}~\bibnamefont
  {Thompson}}\ and\ \bibinfo {author} {\bibfnamefont {R.~C.}\ \bibnamefont
  {Duncan}},\ }\bibfield  {title} {\bibinfo {title} {The soft gamma repeaters
  as very strongly magnetized neutron stars-i. radiative mechanism for
  outbursts},\ }\href@noop {} {\bibfield  {journal} {\bibinfo  {journal}
  {Monthly Notices of the Royal Astronomical Society}\ }\textbf {\bibinfo
  {volume} {275}},\ \bibinfo {pages} {255} (\bibinfo {year}
  {1995})}\BibitemShut {NoStop}%
\bibitem [{\citenamefont {Strohmayer}\ and\ \citenamefont
  {Watts}(2006)}]{strohmayer20062004}%
  \BibitemOpen
  \bibfield  {author} {\bibinfo {author} {\bibfnamefont {T.~E.}\ \bibnamefont
  {Strohmayer}}\ and\ \bibinfo {author} {\bibfnamefont {A.~L.}\ \bibnamefont
  {Watts}},\ }\bibfield  {title} {\bibinfo {title} {The 2004 hyperflare from
  sgr 1806--20: Further evidence for global torsional vibrations},\ }\href@noop
  {} {\bibfield  {journal} {\bibinfo  {journal} {The Astrophysical Journal}\
  }\textbf {\bibinfo {volume} {653}},\ \bibinfo {pages} {593} (\bibinfo {year}
  {2006})}\BibitemShut {NoStop}%
\bibitem [{\citenamefont {Piekarewicz}\ \emph {et~al.}(2014)\citenamefont
  {Piekarewicz}, \citenamefont {Fattoyev},\ and\ \citenamefont
  {Horowitz}}]{piekarewiczPulsarGlitchesCrust2014}%
  \BibitemOpen
  \bibfield  {author} {\bibinfo {author} {\bibfnamefont {J.}~\bibnamefont
  {Piekarewicz}}, \bibinfo {author} {\bibfnamefont {F.~J.}\ \bibnamefont
  {Fattoyev}},\ and\ \bibinfo {author} {\bibfnamefont {C.~J.}\ \bibnamefont
  {Horowitz}},\ }\bibfield  {title} {\bibinfo {title} {Pulsar glitches: {{The}}
  crust may be enough},\ }\href {https://doi.org/10.1103/PhysRevC.90.015803}
  {\bibfield  {journal} {\bibinfo  {journal} {Physical Review C}\ }\textbf
  {\bibinfo {volume} {90}},\ \bibinfo {pages} {015803} (\bibinfo {year}
  {2014})}\BibitemShut {NoStop}%
\bibitem [{\citenamefont {Link}\ \emph {et~al.}(1999)\citenamefont {Link},
  \citenamefont {Epstein},\ and\ \citenamefont {Lattimer}}]{link1999pulsar}%
  \BibitemOpen
  \bibfield  {author} {\bibinfo {author} {\bibfnamefont {B.}~\bibnamefont
  {Link}}, \bibinfo {author} {\bibfnamefont {R.~I.}\ \bibnamefont {Epstein}},\
  and\ \bibinfo {author} {\bibfnamefont {J.~M.}\ \bibnamefont {Lattimer}},\
  }\bibfield  {title} {\bibinfo {title} {Pulsar constraints on neutron star
  structure and equation of state},\ }\href@noop {} {\bibfield  {journal}
  {\bibinfo  {journal} {Physical Review Letters}\ }\textbf {\bibinfo {volume}
  {83}},\ \bibinfo {pages} {3362} (\bibinfo {year} {1999})}\BibitemShut
  {NoStop}%
\bibitem [{\citenamefont {Burrows}\ \emph {et~al.}(2006)\citenamefont
  {Burrows}, \citenamefont {Reddy},\ and\ \citenamefont
  {Thompson}}]{burrows2006neutrino}%
  \BibitemOpen
  \bibfield  {author} {\bibinfo {author} {\bibfnamefont {A.}~\bibnamefont
  {Burrows}}, \bibinfo {author} {\bibfnamefont {S.}~\bibnamefont {Reddy}},\
  and\ \bibinfo {author} {\bibfnamefont {T.~A.}\ \bibnamefont {Thompson}},\
  }\bibfield  {title} {\bibinfo {title} {Neutrino opacities in nuclear
  matter},\ }\href@noop {} {\bibfield  {journal} {\bibinfo  {journal} {Nuclear
  Physics A}\ }\textbf {\bibinfo {volume} {777}},\ \bibinfo {pages} {356}
  (\bibinfo {year} {2006})}\BibitemShut {NoStop}%
\bibitem [{\citenamefont {Brown}(2000)}]{brown2000nuclear}%
  \BibitemOpen
  \bibfield  {author} {\bibinfo {author} {\bibfnamefont {E.~F.}\ \bibnamefont
  {Brown}},\ }\bibfield  {title} {\bibinfo {title} {Nuclear heating and melted
  layers in the inner crust of an accreting neutron star},\ }\href@noop {}
  {\bibfield  {journal} {\bibinfo  {journal} {The Astrophysical Journal}\
  }\textbf {\bibinfo {volume} {531}},\ \bibinfo {pages} {988} (\bibinfo {year}
  {2000})}\BibitemShut {NoStop}%
\bibitem [{\citenamefont {Ferreira}\ and\ \citenamefont
  {Provid{\^e}ncia}(2020)}]{ferreiraEffectCrustNeutron2020a}%
  \BibitemOpen
  \bibfield  {author} {\bibinfo {author} {\bibfnamefont {M.}~\bibnamefont
  {Ferreira}}\ and\ \bibinfo {author} {\bibfnamefont {C.}~\bibnamefont
  {Provid{\^e}ncia}},\ }\bibfield  {title} {\bibinfo {title} {Effect of the
  crust on neutron star empirical relations},\ }\href
  {https://doi.org/10.1103/PhysRevD.102.103003} {\bibfield  {journal} {\bibinfo
   {journal} {Physical Review D}\ }\textbf {\bibinfo {volume} {102}},\ \bibinfo
  {pages} {103003} (\bibinfo {year} {2020})}\BibitemShut {NoStop}%
\bibitem [{\citenamefont {Piekarewicz}\ and\ \citenamefont
  {Fattoyev}(2019)}]{piekarewiczImpactNeutronStar2019}%
  \BibitemOpen
  \bibfield  {author} {\bibinfo {author} {\bibfnamefont {J.}~\bibnamefont
  {Piekarewicz}}\ and\ \bibinfo {author} {\bibfnamefont {F.~J.}\ \bibnamefont
  {Fattoyev}},\ }\bibfield  {title} {\bibinfo {title} {Impact of the neutron
  star crust on the tidal polarizability},\ }\href
  {https://doi.org/10.1103/PhysRevC.99.045802} {\bibfield  {journal} {\bibinfo
  {journal} {Physical Review C}\ }\textbf {\bibinfo {volume} {99}},\ \bibinfo
  {pages} {045802} (\bibinfo {year} {2019})}\BibitemShut {NoStop}%
\bibitem [{\citenamefont {Perot}\ \emph {et~al.}(2020)\citenamefont {Perot},
  \citenamefont {Chamel},\ and\ \citenamefont
  {Sourie}}]{perotRoleCrustTidal2020}%
  \BibitemOpen
  \bibfield  {author} {\bibinfo {author} {\bibfnamefont {L.}~\bibnamefont
  {Perot}}, \bibinfo {author} {\bibfnamefont {N.}~\bibnamefont {Chamel}},\ and\
  \bibinfo {author} {\bibfnamefont {A.}~\bibnamefont {Sourie}},\ }\bibfield
  {title} {\bibinfo {title} {Role of the crust in the tidal deformability of a
  neutron star within a unified treatment of dense matter},\ }\href
  {https://doi.org/10.1103/PhysRevC.101.015806} {\bibfield  {journal} {\bibinfo
   {journal} {Physical Review C}\ }\textbf {\bibinfo {volume} {101}},\ \bibinfo
  {pages} {015806} (\bibinfo {year} {2020})}\BibitemShut {NoStop}%
\bibitem [{\citenamefont {Baym}\ \emph {et~al.}(1971)\citenamefont {Baym},
  \citenamefont {Pethick},\ and\ \citenamefont {Sutherland}}]{baym1971ground}%
  \BibitemOpen
  \bibfield  {author} {\bibinfo {author} {\bibfnamefont {G.}~\bibnamefont
  {Baym}}, \bibinfo {author} {\bibfnamefont {C.}~\bibnamefont {Pethick}},\ and\
  \bibinfo {author} {\bibfnamefont {P.}~\bibnamefont {Sutherland}},\ }\bibfield
   {title} {\bibinfo {title} {The ground state of matter at high densities:
  equation of state and stellar models},\ }\href@noop {} {\bibfield  {journal}
  {\bibinfo  {journal} {The Astrophysical Journal}\ }\textbf {\bibinfo {volume}
  {170}},\ \bibinfo {pages} {299} (\bibinfo {year} {1971})}\BibitemShut
  {NoStop}%
\bibitem [{\citenamefont {Pastore}\ \emph {et~al.}(2020)\citenamefont
  {Pastore}, \citenamefont {Neill}, \citenamefont {Powell}, \citenamefont
  {Medler},\ and\ \citenamefont {Barton}}]{pastore2020impact}%
  \BibitemOpen
  \bibfield  {author} {\bibinfo {author} {\bibfnamefont {A.}~\bibnamefont
  {Pastore}}, \bibinfo {author} {\bibfnamefont {D.}~\bibnamefont {Neill}},
  \bibinfo {author} {\bibfnamefont {H.}~\bibnamefont {Powell}}, \bibinfo
  {author} {\bibfnamefont {K.}~\bibnamefont {Medler}},\ and\ \bibinfo {author}
  {\bibfnamefont {C.}~\bibnamefont {Barton}},\ }\bibfield  {title} {\bibinfo
  {title} {Impact of statistical uncertainties on the composition of the outer
  crust of a neutron star},\ }\href@noop {} {\bibfield  {journal} {\bibinfo
  {journal} {Physical Review C}\ }\textbf {\bibinfo {volume} {101}},\ \bibinfo
  {pages} {035804} (\bibinfo {year} {2020})}\BibitemShut {NoStop}%
\bibitem [{\citenamefont {Fantina}\ \emph {et~al.}(2020)\citenamefont
  {Fantina}, \citenamefont {De~Ridder}, \citenamefont {Chamel},\ and\
  \citenamefont {Gulminelli}}]{fantina2020crystallization}%
  \BibitemOpen
  \bibfield  {author} {\bibinfo {author} {\bibfnamefont {A.}~\bibnamefont
  {Fantina}}, \bibinfo {author} {\bibfnamefont {S.}~\bibnamefont {De~Ridder}},
  \bibinfo {author} {\bibfnamefont {N.}~\bibnamefont {Chamel}},\ and\ \bibinfo
  {author} {\bibfnamefont {F.}~\bibnamefont {Gulminelli}},\ }\bibfield  {title}
  {\bibinfo {title} {Crystallization of the outer crust of a non-accreting
  neutron star},\ }\href@noop {} {\bibfield  {journal} {\bibinfo  {journal}
  {Astronomy \& Astrophysics}\ }\textbf {\bibinfo {volume} {633}},\ \bibinfo
  {pages} {A149} (\bibinfo {year} {2020})}\BibitemShut {NoStop}%
\bibitem [{\citenamefont {Negele}\ and\ \citenamefont
  {Vautherin}(1973)}]{negele1973neutron}%
  \BibitemOpen
  \bibfield  {author} {\bibinfo {author} {\bibfnamefont {J.~W.}\ \bibnamefont
  {Negele}}\ and\ \bibinfo {author} {\bibfnamefont {D.}~\bibnamefont
  {Vautherin}},\ }\bibfield  {title} {\bibinfo {title} {Neutron star matter at
  sub-nuclear densities},\ }\href@noop {} {\bibfield  {journal} {\bibinfo
  {journal} {Nuclear Physics A}\ }\textbf {\bibinfo {volume} {207}},\ \bibinfo
  {pages} {298} (\bibinfo {year} {1973})}\BibitemShut {NoStop}%
\bibitem [{\citenamefont {Avogadro}\ \emph {et~al.}(2007)\citenamefont
  {Avogadro}, \citenamefont {Barranco}, \citenamefont {Broglia},\ and\
  \citenamefont {Vigezzi}}]{avogadro2007quantum}%
  \BibitemOpen
  \bibfield  {author} {\bibinfo {author} {\bibfnamefont {P.}~\bibnamefont
  {Avogadro}}, \bibinfo {author} {\bibfnamefont {F.}~\bibnamefont {Barranco}},
  \bibinfo {author} {\bibfnamefont {R.}~\bibnamefont {Broglia}},\ and\ \bibinfo
  {author} {\bibfnamefont {E.}~\bibnamefont {Vigezzi}},\ }\bibfield  {title}
  {\bibinfo {title} {Quantum calculation of vortices in the inner crust of
  neutron stars},\ }\href@noop {} {\bibfield  {journal} {\bibinfo  {journal}
  {Physical Review C}\ }\textbf {\bibinfo {volume} {75}},\ \bibinfo {pages}
  {012805} (\bibinfo {year} {2007})}\BibitemShut {NoStop}%
\bibitem [{\citenamefont {Pearson}\ \emph {et~al.}(2012)\citenamefont
  {Pearson}, \citenamefont {Chamel}, \citenamefont {Goriely},\ and\
  \citenamefont {Ducoin}}]{pearsonInnerCrustNeutron2012}%
  \BibitemOpen
  \bibfield  {author} {\bibinfo {author} {\bibfnamefont {J.~M.}\ \bibnamefont
  {Pearson}}, \bibinfo {author} {\bibfnamefont {N.}~\bibnamefont {Chamel}},
  \bibinfo {author} {\bibfnamefont {S.}~\bibnamefont {Goriely}},\ and\ \bibinfo
  {author} {\bibfnamefont {C.}~\bibnamefont {Ducoin}},\ }\bibfield  {title}
  {\bibinfo {title} {Inner crust of neutron stars with mass-fitted {{Skyrme}}
  functionals},\ }\href {https://doi.org/10.1103/PhysRevC.85.065803} {\bibfield
   {journal} {\bibinfo  {journal} {Physical Review C}\ }\textbf {\bibinfo
  {volume} {85}},\ \bibinfo {pages} {065803} (\bibinfo {year}
  {2012})}\BibitemShut {NoStop}%
\bibitem [{\citenamefont {Watanabe}\ and\ \citenamefont
  {Iida}(2003)}]{watanabe2003electron}%
  \BibitemOpen
  \bibfield  {author} {\bibinfo {author} {\bibfnamefont {G.}~\bibnamefont
  {Watanabe}}\ and\ \bibinfo {author} {\bibfnamefont {K.}~\bibnamefont
  {Iida}},\ }\bibfield  {title} {\bibinfo {title} {Electron screening in the
  liquid-gas mixed phases of nuclear matter},\ }\href@noop {} {\bibfield
  {journal} {\bibinfo  {journal} {Physical Review C}\ }\textbf {\bibinfo
  {volume} {68}},\ \bibinfo {pages} {045801} (\bibinfo {year}
  {2003})}\BibitemShut {NoStop}%
\bibitem [{\citenamefont {Douchin}\ and\ \citenamefont
  {Haensel}(2000)}]{douchin2000inner}%
  \BibitemOpen
  \bibfield  {author} {\bibinfo {author} {\bibfnamefont {F.}~\bibnamefont
  {Douchin}}\ and\ \bibinfo {author} {\bibfnamefont {P.}~\bibnamefont
  {Haensel}},\ }\bibfield  {title} {\bibinfo {title} {Inner edge of
  neutron-star crust with sly effective nucleon-nucleon interactions},\
  }\href@noop {} {\bibfield  {journal} {\bibinfo  {journal} {Physics Letters
  B}\ }\textbf {\bibinfo {volume} {485}},\ \bibinfo {pages} {107} (\bibinfo
  {year} {2000})}\BibitemShut {NoStop}%
\bibitem [{\citenamefont {Oyamatsu}\ and\ \citenamefont
  {Yamada}(1994)}]{oyamatsu1994shell}%
  \BibitemOpen
  \bibfield  {author} {\bibinfo {author} {\bibfnamefont {K.}~\bibnamefont
  {Oyamatsu}}\ and\ \bibinfo {author} {\bibfnamefont {M.}~\bibnamefont
  {Yamada}},\ }\bibfield  {title} {\bibinfo {title} {Shell energies of
  non-spherical nuclei in the inner crust of a neutron star},\ }\href@noop {}
  {\bibfield  {journal} {\bibinfo  {journal} {Nuclear Physics A}\ }\textbf
  {\bibinfo {volume} {578}},\ \bibinfo {pages} {181} (\bibinfo {year}
  {1994})}\BibitemShut {NoStop}%
\bibitem [{\citenamefont {Onsi}\ \emph {et~al.}(1997)\citenamefont {Onsi},
  \citenamefont {Przysiezniak},\ and\ \citenamefont
  {Pearson}}]{onsiEquationStateStellar1997a}%
  \BibitemOpen
  \bibfield  {author} {\bibinfo {author} {\bibfnamefont {M.}~\bibnamefont
  {Onsi}}, \bibinfo {author} {\bibfnamefont {H.}~\bibnamefont {Przysiezniak}},\
  and\ \bibinfo {author} {\bibfnamefont {J.~M.}\ \bibnamefont {Pearson}},\
  }\bibfield  {title} {\bibinfo {title} {Equation of state of stellar nuclear
  matter in the temperature-dependent extended {{Thomas}}-{{Fermi}}
  formalism},\ }\href {https://doi.org/10.1103/PhysRevC.55.3139} {\bibfield
  {journal} {\bibinfo  {journal} {Physical Review C}\ }\textbf {\bibinfo
  {volume} {55}},\ \bibinfo {pages} {3139} (\bibinfo {year}
  {1997})}\BibitemShut {NoStop}%
\bibitem [{\citenamefont {Onsi}\ \emph {et~al.}(2008)\citenamefont {Onsi},
  \citenamefont {Dutta}, \citenamefont {Chatri}, \citenamefont {Goriely},
  \citenamefont {Chamel},\ and\ \citenamefont {Pearson}}]{onsi2008semi}%
  \BibitemOpen
  \bibfield  {author} {\bibinfo {author} {\bibfnamefont {M.}~\bibnamefont
  {Onsi}}, \bibinfo {author} {\bibfnamefont {A.}~\bibnamefont {Dutta}},
  \bibinfo {author} {\bibfnamefont {H.}~\bibnamefont {Chatri}}, \bibinfo
  {author} {\bibfnamefont {S.}~\bibnamefont {Goriely}}, \bibinfo {author}
  {\bibfnamefont {N.}~\bibnamefont {Chamel}},\ and\ \bibinfo {author}
  {\bibfnamefont {J.}~\bibnamefont {Pearson}},\ }\bibfield  {title} {\bibinfo
  {title} {Semi-classical equation of state and specific-heat expressions with
  proton shell corrections for the inner crust of a neutron star},\ }\href@noop
  {} {\bibfield  {journal} {\bibinfo  {journal} {Physical Review C}\ }\textbf
  {\bibinfo {volume} {77}},\ \bibinfo {pages} {065805} (\bibinfo {year}
  {2008})}\BibitemShut {NoStop}%
\bibitem [{\citenamefont {Martin}\ and\ \citenamefont
  {Urban}(2015)}]{martinLiquidgasCoexistenceEnergy2015}%
  \BibitemOpen
  \bibfield  {author} {\bibinfo {author} {\bibfnamefont {N.}~\bibnamefont
  {Martin}}\ and\ \bibinfo {author} {\bibfnamefont {M.}~\bibnamefont {Urban}},\
  }\bibfield  {title} {\bibinfo {title} {Liquid-gas coexistence versus energy
  minimization with respect to the density profile in the inhomogeneous inner
  crust of neutron stars},\ }\href {https://doi.org/10.1103/PhysRevC.92.015803}
  {\bibfield  {journal} {\bibinfo  {journal} {Physical Review C}\ }\textbf
  {\bibinfo {volume} {92}},\ \bibinfo {pages} {015803} (\bibinfo {year}
  {2015})}\BibitemShut {NoStop}%
\bibitem [{\citenamefont {Sharma}\ \emph {et~al.}(2015)\citenamefont {Sharma},
  \citenamefont {Centelles}, \citenamefont {Vi{\~n}as}, \citenamefont {Baldo},\
  and\ \citenamefont {Burgio}}]{sharmaUnifiedEquationState2015}%
  \BibitemOpen
  \bibfield  {author} {\bibinfo {author} {\bibfnamefont {B.~K.}\ \bibnamefont
  {Sharma}}, \bibinfo {author} {\bibfnamefont {M.}~\bibnamefont {Centelles}},
  \bibinfo {author} {\bibfnamefont {X.}~\bibnamefont {Vi{\~n}as}}, \bibinfo
  {author} {\bibfnamefont {M.}~\bibnamefont {Baldo}},\ and\ \bibinfo {author}
  {\bibfnamefont {G.~F.}\ \bibnamefont {Burgio}},\ }\bibfield  {title}
  {\bibinfo {title} {Unified equation of state for neutron stars on a
  microscopic basis},\ }\href {https://doi.org/10.1051/0004-6361/201526642}
  {\bibfield  {journal} {\bibinfo  {journal} {Astronomy \& Astrophysics}\
  }\textbf {\bibinfo {volume} {584}},\ \bibinfo {pages} {A103} (\bibinfo {year}
  {2015})}\BibitemShut {NoStop}%
\bibitem [{\citenamefont {Lim}\ and\ \citenamefont
  {Holt}(2017)}]{limStructureNeutronStar2017}%
  \BibitemOpen
  \bibfield  {author} {\bibinfo {author} {\bibfnamefont {Y.}~\bibnamefont
  {Lim}}\ and\ \bibinfo {author} {\bibfnamefont {J.~W.}\ \bibnamefont {Holt}},\
  }\bibfield  {title} {\bibinfo {title} {Structure of neutron star crusts from
  new {{Skyrme}} effective interactions constrained by chiral effective field
  theory},\ }\href {https://doi.org/10.1103/PhysRevC.95.065805} {\bibfield
  {journal} {\bibinfo  {journal} {Physical Review C}\ }\textbf {\bibinfo
  {volume} {95}},\ \bibinfo {pages} {065805} (\bibinfo {year}
  {2017})}\BibitemShut {NoStop}%
\bibitem [{\citenamefont {Grill}\ \emph {et~al.}(2011)\citenamefont {Grill},
  \citenamefont {Margueron},\ and\ \citenamefont
  {Sandulescu}}]{grill2011cluster}%
  \BibitemOpen
  \bibfield  {author} {\bibinfo {author} {\bibfnamefont {F.}~\bibnamefont
  {Grill}}, \bibinfo {author} {\bibfnamefont {J.}~\bibnamefont {Margueron}},\
  and\ \bibinfo {author} {\bibfnamefont {N.}~\bibnamefont {Sandulescu}},\
  }\bibfield  {title} {\bibinfo {title} {Cluster structure of the inner crust
  of neutron stars in the hartree-fock-bogoliubov approach},\ }\href@noop {}
  {\bibfield  {journal} {\bibinfo  {journal} {Physical Review C}\ }\textbf
  {\bibinfo {volume} {84}},\ \bibinfo {pages} {065801} (\bibinfo {year}
  {2011})}\BibitemShut {NoStop}%
\bibitem [{\citenamefont {Pastore}\ \emph {et~al.}(2011)\citenamefont
  {Pastore}, \citenamefont {Baroni},\ and\ \citenamefont
  {Losa}}]{pastore2011superfluid}%
  \BibitemOpen
  \bibfield  {author} {\bibinfo {author} {\bibfnamefont {A.}~\bibnamefont
  {Pastore}}, \bibinfo {author} {\bibfnamefont {S.}~\bibnamefont {Baroni}},\
  and\ \bibinfo {author} {\bibfnamefont {C.}~\bibnamefont {Losa}},\ }\bibfield
  {title} {\bibinfo {title} {Superfluid properties of the inner crust of
  neutron stars},\ }\href@noop {} {\bibfield  {journal} {\bibinfo  {journal}
  {Physical Review C}\ }\textbf {\bibinfo {volume} {84}},\ \bibinfo {pages}
  {065807} (\bibinfo {year} {2011})}\BibitemShut {NoStop}%
\bibitem [{\citenamefont {Pastore}\ \emph {et~al.}(2017)\citenamefont
  {Pastore}, \citenamefont {Shelley}, \citenamefont {Baroni},\ and\
  \citenamefont {Diget}}]{pastore2017new}%
  \BibitemOpen
  \bibfield  {author} {\bibinfo {author} {\bibfnamefont {A.}~\bibnamefont
  {Pastore}}, \bibinfo {author} {\bibfnamefont {M.}~\bibnamefont {Shelley}},
  \bibinfo {author} {\bibfnamefont {S.}~\bibnamefont {Baroni}},\ and\ \bibinfo
  {author} {\bibfnamefont {C.}~\bibnamefont {Diget}},\ }\bibfield  {title}
  {\bibinfo {title} {A new statistical method for the structure of the inner
  crust of neutron stars},\ }\href@noop {} {\bibfield  {journal} {\bibinfo
  {journal} {Journal of Physics G: Nuclear and Particle Physics}\ }\textbf
  {\bibinfo {volume} {44}},\ \bibinfo {pages} {094003} (\bibinfo {year}
  {2017})}\BibitemShut {NoStop}%
\bibitem [{\citenamefont {Skyrme}(1956)}]{skyrme1956cvii}%
  \BibitemOpen
  \bibfield  {author} {\bibinfo {author} {\bibfnamefont {T.}~\bibnamefont
  {Skyrme}},\ }\bibfield  {title} {\bibinfo {title} {Cvii. the nuclear
  surface},\ }\href@noop {} {\bibfield  {journal} {\bibinfo  {journal}
  {Philosophical Magazine}\ }\textbf {\bibinfo {volume} {1}},\ \bibinfo {pages}
  {1043} (\bibinfo {year} {1956})}\BibitemShut {NoStop}%
\bibitem [{\citenamefont {Reinhard}\ and\ \citenamefont
  {Bender}(2004)}]{Book:Reinhard2004}%
  \BibitemOpen
  \bibfield  {author} {\bibinfo {author} {\bibfnamefont {P.-G.}\ \bibnamefont
  {Reinhard}}\ and\ \bibinfo {author} {\bibfnamefont {M.}~\bibnamefont
  {Bender}},\ }\bibfield  {title} {\bibinfo {title} {{Mean Field : Relativistic
  versus Non-relativistic}},\ }in\ \href@noop {} {\emph {\bibinfo {booktitle}
  {Lectures Notes in Physics "Extended Density Functionals in Nuclear Structure
  Physics"}}},\ Vol.\ \bibinfo {volume} {268}\ (\bibinfo  {publisher}
  {Springer},\ \bibinfo {year} {2004})\ pp.\ \bibinfo {pages}
  {249--268}\BibitemShut {NoStop}%
\bibitem [{\citenamefont {Kortelainen}\ \emph {et~al.}(2014)\citenamefont
  {Kortelainen}, \citenamefont {McDonnell}, \citenamefont {Nazarewicz},
  \citenamefont {Olsen}, \citenamefont {Reinhard}, \citenamefont {Sarich},
  \citenamefont {Schunck}, \citenamefont {Wild}, \citenamefont {Davesne},
  \citenamefont {Erler} \emph {et~al.}}]{kortelainen2014nuclear}%
  \BibitemOpen
  \bibfield  {author} {\bibinfo {author} {\bibfnamefont {M.}~\bibnamefont
  {Kortelainen}}, \bibinfo {author} {\bibfnamefont {J.}~\bibnamefont
  {McDonnell}}, \bibinfo {author} {\bibfnamefont {W.}~\bibnamefont
  {Nazarewicz}}, \bibinfo {author} {\bibfnamefont {E.}~\bibnamefont {Olsen}},
  \bibinfo {author} {\bibfnamefont {P.-G.}\ \bibnamefont {Reinhard}}, \bibinfo
  {author} {\bibfnamefont {J.}~\bibnamefont {Sarich}}, \bibinfo {author}
  {\bibfnamefont {N.}~\bibnamefont {Schunck}}, \bibinfo {author} {\bibfnamefont
  {S.}~\bibnamefont {Wild}}, \bibinfo {author} {\bibfnamefont {D.}~\bibnamefont
  {Davesne}}, \bibinfo {author} {\bibfnamefont {J.}~\bibnamefont {Erler}},
  \emph {et~al.},\ }\bibfield  {title} {\bibinfo {title} {Nuclear energy
  density optimization: Shell structure},\ }\href@noop {} {\bibfield  {journal}
  {\bibinfo  {journal} {Physical Review C}\ }\textbf {\bibinfo {volume} {89}},\
  \bibinfo {pages} {054314} (\bibinfo {year} {2014})}\BibitemShut {NoStop}%
\bibitem [{\citenamefont {Baldo}\ \emph {et~al.}(2006)\citenamefont {Baldo},
  \citenamefont {Saperstein},\ and\ \citenamefont
  {Tolokonnikov}}]{baldo2006role}%
  \BibitemOpen
  \bibfield  {author} {\bibinfo {author} {\bibfnamefont {M.}~\bibnamefont
  {Baldo}}, \bibinfo {author} {\bibfnamefont {E.}~\bibnamefont {Saperstein}},\
  and\ \bibinfo {author} {\bibfnamefont {S.}~\bibnamefont {Tolokonnikov}},\
  }\bibfield  {title} {\bibinfo {title} {The role of the boundary conditions in
  the wigner--seitz approximation applied to the neutron star inner crust},\
  }\href@noop {} {\bibfield  {journal} {\bibinfo  {journal} {Nuclear Physics
  A}\ }\textbf {\bibinfo {volume} {775}},\ \bibinfo {pages} {235} (\bibinfo
  {year} {2006})}\BibitemShut {NoStop}%
\bibitem [{\citenamefont {Margueron}\ \emph {et~al.}(2008)\citenamefont
  {Margueron}, \citenamefont {Van~Giai},\ and\ \citenamefont
  {Sandulescu}}]{margueronEquationStateInner2008}%
  \BibitemOpen
  \bibfield  {author} {\bibinfo {author} {\bibfnamefont {J.}~\bibnamefont
  {Margueron}}, \bibinfo {author} {\bibfnamefont {N.}~\bibnamefont
  {Van~Giai}},\ and\ \bibinfo {author} {\bibfnamefont {N.}~\bibnamefont
  {Sandulescu}},\ }\bibfield  {title} {\bibinfo {title} {Equation of state in
  the inner crust of neutron stars: Discussion of the unbound neutrons
  states},\ }in\ \href {https://doi.org/10.1142/9789812797049_0059} {\emph
  {\bibinfo {booktitle} {Exotic {{States}} of {{Nuclear Matter}}}}}\ (\bibinfo
  {publisher} {World Scientific, Singapore},\ \bibinfo {year} {2008})\ pp.\
  \bibinfo {pages} {362--369}\BibitemShut {NoStop}%
\bibitem [{\citenamefont {Chamel}(2012)}]{chamel2012neutron}%
  \BibitemOpen
  \bibfield  {author} {\bibinfo {author} {\bibfnamefont {N.}~\bibnamefont
  {Chamel}},\ }\bibfield  {title} {\bibinfo {title} {Neutron conduction in the
  inner crust of a neutron star in the framework of the band theory of
  solids},\ }\href@noop {} {\bibfield  {journal} {\bibinfo  {journal} {Physical
  Review C}\ }\textbf {\bibinfo {volume} {85}},\ \bibinfo {pages} {035801}
  (\bibinfo {year} {2012})}\BibitemShut {NoStop}%
\bibitem [{\citenamefont {Jin}\ \emph {et~al.}(2017)\citenamefont {Jin},
  \citenamefont {Bulgac}, \citenamefont {Roche},\ and\ \citenamefont
  {Wlaz{\l}owski}}]{jin2017coordinate}%
  \BibitemOpen
  \bibfield  {author} {\bibinfo {author} {\bibfnamefont {S.}~\bibnamefont
  {Jin}}, \bibinfo {author} {\bibfnamefont {A.}~\bibnamefont {Bulgac}},
  \bibinfo {author} {\bibfnamefont {K.}~\bibnamefont {Roche}},\ and\ \bibinfo
  {author} {\bibfnamefont {G.}~\bibnamefont {Wlaz{\l}owski}},\ }\bibfield
  {title} {\bibinfo {title} {Coordinate-space solver for superfluid
  many-fermion systems with the shifted conjugate-orthogonal conjugate-gradient
  method},\ }\href@noop {} {\bibfield  {journal} {\bibinfo  {journal} {Physical
  Review C}\ }\textbf {\bibinfo {volume} {95}},\ \bibinfo {pages} {044302}
  (\bibinfo {year} {2017})}\BibitemShut {NoStop}%
\bibitem [{\citenamefont {Mondal}\ \emph {et~al.}(2020)\citenamefont {Mondal},
  \citenamefont {Vi{\~n}as}, \citenamefont {Centelles},\ and\ \citenamefont
  {De}}]{mondalStructureCompositionInner2020}%
  \BibitemOpen
  \bibfield  {author} {\bibinfo {author} {\bibfnamefont {C.}~\bibnamefont
  {Mondal}}, \bibinfo {author} {\bibfnamefont {X.}~\bibnamefont {Vi{\~n}as}},
  \bibinfo {author} {\bibfnamefont {M.}~\bibnamefont {Centelles}},\ and\
  \bibinfo {author} {\bibfnamefont {J.~N.}\ \bibnamefont {De}},\ }\bibfield
  {title} {\bibinfo {title} {Structure and composition of the inner crust of
  neutron stars from {{Gogny}} interactions},\ }\href
  {https://doi.org/10.1103/PhysRevC.102.015802} {\bibfield  {journal} {\bibinfo
   {journal} {Physical Review C}\ }\textbf {\bibinfo {volume} {102}},\ \bibinfo
  {pages} {015802} (\bibinfo {year} {2020})}\BibitemShut {NoStop}%
\bibitem [{\citenamefont {Pearson}\ \emph {et~al.}(2015)\citenamefont
  {Pearson}, \citenamefont {Chamel}, \citenamefont {Pastore},\ and\
  \citenamefont {Goriely}}]{pearson2015role}%
  \BibitemOpen
  \bibfield  {author} {\bibinfo {author} {\bibfnamefont {J.}~\bibnamefont
  {Pearson}}, \bibinfo {author} {\bibfnamefont {N.}~\bibnamefont {Chamel}},
  \bibinfo {author} {\bibfnamefont {A.}~\bibnamefont {Pastore}},\ and\ \bibinfo
  {author} {\bibfnamefont {S.}~\bibnamefont {Goriely}},\ }\bibfield  {title}
  {\bibinfo {title} {Role of proton pairing in a semimicroscopic treatment of
  the inner crust of neutron stars},\ }\href@noop {} {\bibfield  {journal}
  {\bibinfo  {journal} {Physical Review C}\ }\textbf {\bibinfo {volume} {91}},\
  \bibinfo {pages} {018801} (\bibinfo {year} {2015})}\BibitemShut {NoStop}%
\bibitem [{\citenamefont {Shelley}\ and\ \citenamefont
  {Pastore}(2020{\natexlab{a}})}]{shelley2020accurately}%
  \BibitemOpen
  \bibfield  {author} {\bibinfo {author} {\bibfnamefont {M.}~\bibnamefont
  {Shelley}}\ and\ \bibinfo {author} {\bibfnamefont {A.}~\bibnamefont
  {Pastore}},\ }\bibfield  {title} {\bibinfo {title} {How accurately can the
  extended thomas-fermi method describe the inner crust of a neutron star?},\
  }\href@noop {} {\bibfield  {journal} {\bibinfo  {journal} {Journal of
  Physics: Conference Series}\ }\textbf {\bibinfo {volume} {2020}},\ \bibinfo
  {pages} {012037} (\bibinfo {year} {2020}{\natexlab{a}})}\BibitemShut
  {NoStop}%
\bibitem [{\citenamefont {Shelley}\ and\ \citenamefont
  {Pastore}(2020{\natexlab{b}})}]{universe6110206}%
  \BibitemOpen
  \bibfield  {author} {\bibinfo {author} {\bibfnamefont {M.}~\bibnamefont
  {Shelley}}\ and\ \bibinfo {author} {\bibfnamefont {A.}~\bibnamefont
  {Pastore}},\ }\bibfield  {title} {\bibinfo {title} {Comparison between the
  thomas-fermi and hartree-fock-bogoliubov methods in the inner crust of a
  neutron star: The role of pairing correlations},\ }\href@noop {} {\bibfield
  {journal} {\bibinfo  {journal} {Universe}\ }\textbf {\bibinfo {volume} {6}},\
  \bibinfo {pages} {206} (\bibinfo {year} {2020}{\natexlab{b}})}\BibitemShut
  {NoStop}%
\bibitem [{\citenamefont {Pearson}\ \emph {et~al.}(2018)\citenamefont
  {Pearson}, \citenamefont {Chamel}, \citenamefont {Potekhin}, \citenamefont
  {Fantina}, \citenamefont {Ducoin}, \citenamefont {Dutta},\ and\ \citenamefont
  {Goriely}}]{pearsonUnifiedEquationsState2018}%
  \BibitemOpen
  \bibfield  {author} {\bibinfo {author} {\bibfnamefont {J.~M.}\ \bibnamefont
  {Pearson}}, \bibinfo {author} {\bibfnamefont {N.}~\bibnamefont {Chamel}},
  \bibinfo {author} {\bibfnamefont {A.~Y.}\ \bibnamefont {Potekhin}}, \bibinfo
  {author} {\bibfnamefont {A.~F.}\ \bibnamefont {Fantina}}, \bibinfo {author}
  {\bibfnamefont {C.}~\bibnamefont {Ducoin}}, \bibinfo {author} {\bibfnamefont
  {A.~K.}\ \bibnamefont {Dutta}},\ and\ \bibinfo {author} {\bibfnamefont
  {S.}~\bibnamefont {Goriely}},\ }\bibfield  {title} {\bibinfo {title} {Unified
  equations of state for cold non-accreting neutron stars with
  {{Brussels}}\textendash{{Montreal}} functionals \textendash{} {{I}}. {{Role}}
  of symmetry energy},\ }\href {https://doi.org/10.1093/mnras/sty2413}
  {\bibfield  {journal} {\bibinfo  {journal} {Monthly Notices of the Royal
  Astronomical Society}\ }\textbf {\bibinfo {volume} {481}},\ \bibinfo {pages}
  {2994} (\bibinfo {year} {2018})}\BibitemShut {NoStop}%
\bibitem [{\citenamefont {Pearson}\ \emph {et~al.}(2019)\citenamefont
  {Pearson}, \citenamefont {Chamel}, \citenamefont {Potekhin}, \citenamefont
  {Fantina}, \citenamefont {Ducoin}, \citenamefont {Dutta},\ and\ \citenamefont
  {Goriely}}]{pearsonErratumUnifiedEquations2019}%
  \BibitemOpen
  \bibfield  {author} {\bibinfo {author} {\bibfnamefont {J.~M.}\ \bibnamefont
  {Pearson}}, \bibinfo {author} {\bibfnamefont {N.}~\bibnamefont {Chamel}},
  \bibinfo {author} {\bibfnamefont {A.~Y.}\ \bibnamefont {Potekhin}}, \bibinfo
  {author} {\bibfnamefont {A.~F.}\ \bibnamefont {Fantina}}, \bibinfo {author}
  {\bibfnamefont {C.}~\bibnamefont {Ducoin}}, \bibinfo {author} {\bibfnamefont
  {A.~K.}\ \bibnamefont {Dutta}},\ and\ \bibinfo {author} {\bibfnamefont
  {S.}~\bibnamefont {Goriely}},\ }\bibfield  {title} {\bibinfo {title}
  {Erratum: {{Unified}} equations of state for cold non-accreting neutron stars
  with {{Brussels}}-{{Montreal}} functionals. {{I}}. {{Role}} of symmetry
  energy},\ }\href {https://doi.org/10.1093/mnras/stz800} {\bibfield  {journal}
  {\bibinfo  {journal} {Monthly Notices of the Royal Astronomical Society}\
  }\textbf {\bibinfo {volume} {486}},\ \bibinfo {pages} {768} (\bibinfo {year}
  {2019})}\BibitemShut {NoStop}%
\bibitem [{\citenamefont {Pearson}\ \emph {et~al.}(2020)\citenamefont
  {Pearson}, \citenamefont {Chamel},\ and\ \citenamefont
  {Potekhin}}]{pearson2020unified}%
  \BibitemOpen
  \bibfield  {author} {\bibinfo {author} {\bibfnamefont {J.~M.}\ \bibnamefont
  {Pearson}}, \bibinfo {author} {\bibfnamefont {N.}~\bibnamefont {Chamel}},\
  and\ \bibinfo {author} {\bibfnamefont {A.}~\bibnamefont {Potekhin}},\
  }\bibfield  {title} {\bibinfo {title} {Unified equations of state for cold
  nonaccreting neutron stars with brussels-montreal functionals. ii. pasta
  phases in semiclassical approximation},\ }\href@noop {} {\bibfield  {journal}
  {\bibinfo  {journal} {Physical Review C}\ }\textbf {\bibinfo {volume}
  {101}},\ \bibinfo {pages} {015802} (\bibinfo {year} {2020})}\BibitemShut
  {NoStop}%
\bibitem [{\citenamefont {Dutra}\ \emph {et~al.}(2012)\citenamefont {Dutra},
  \citenamefont {Louren{\c{c}}o}, \citenamefont {Martins}, \citenamefont
  {Delfino}, \citenamefont {Stone},\ and\ \citenamefont
  {Stevenson}}]{dutra2012skyrme}%
  \BibitemOpen
  \bibfield  {author} {\bibinfo {author} {\bibfnamefont {M.}~\bibnamefont
  {Dutra}}, \bibinfo {author} {\bibfnamefont {O.}~\bibnamefont
  {Louren{\c{c}}o}}, \bibinfo {author} {\bibfnamefont {J.~S.}\ \bibnamefont
  {Martins}}, \bibinfo {author} {\bibfnamefont {A.}~\bibnamefont {Delfino}},
  \bibinfo {author} {\bibfnamefont {J.~R.}\ \bibnamefont {Stone}},\ and\
  \bibinfo {author} {\bibfnamefont {P.}~\bibnamefont {Stevenson}},\ }\bibfield
  {title} {\bibinfo {title} {Skyrme interaction and nuclear matter
  constraints},\ }\href@noop {} {\bibfield  {journal} {\bibinfo  {journal}
  {Physical Review C}\ }\textbf {\bibinfo {volume} {85}},\ \bibinfo {pages}
  {035201} (\bibinfo {year} {2012})}\BibitemShut {NoStop}%
\bibitem [{\citenamefont {Burrello}\ \emph {et~al.}(2015)\citenamefont
  {Burrello}, \citenamefont {Gulminelli}, \citenamefont {Aymard}, \citenamefont
  {Colonna},\ and\ \citenamefont {Raduta}}]{burrello2015heat}%
  \BibitemOpen
  \bibfield  {author} {\bibinfo {author} {\bibfnamefont {S.}~\bibnamefont
  {Burrello}}, \bibinfo {author} {\bibfnamefont {F.}~\bibnamefont
  {Gulminelli}}, \bibinfo {author} {\bibfnamefont {F.}~\bibnamefont {Aymard}},
  \bibinfo {author} {\bibfnamefont {M.}~\bibnamefont {Colonna}},\ and\ \bibinfo
  {author} {\bibfnamefont {A.~R.}\ \bibnamefont {Raduta}},\ }\bibfield  {title}
  {\bibinfo {title} {Heat capacity of the neutron star inner crust within an
  extended nuclear statistical equilibrium model},\ }\href@noop {} {\bibfield
  {journal} {\bibinfo  {journal} {Physical Review C}\ }\textbf {\bibinfo
  {volume} {92}},\ \bibinfo {pages} {055804} (\bibinfo {year}
  {2015})}\BibitemShut {NoStop}%
\bibitem [{\citenamefont {Pastore}(2012)}]{pastore2012superfluid}%
  \BibitemOpen
  \bibfield  {author} {\bibinfo {author} {\bibfnamefont {A.}~\bibnamefont
  {Pastore}},\ }\bibfield  {title} {\bibinfo {title} {Superfluid properties of
  the inner crust of neutron stars. ii. wigner-seitz cells at finite
  temperature},\ }\href@noop {} {\bibfield  {journal} {\bibinfo  {journal}
  {Physical Review C}\ }\textbf {\bibinfo {volume} {86}},\ \bibinfo {pages}
  {065802} (\bibinfo {year} {2012})}\BibitemShut {NoStop}%
\bibitem [{\citenamefont {Shapiro}\ and\ \citenamefont
  {Teukolsky}(2008)}]{shapiro2008black}%
  \BibitemOpen
  \bibfield  {author} {\bibinfo {author} {\bibfnamefont {S.~L.}\ \bibnamefont
  {Shapiro}}\ and\ \bibinfo {author} {\bibfnamefont {S.~A.}\ \bibnamefont
  {Teukolsky}},\ }\href@noop {} {\emph {\bibinfo {title} {Black holes, white
  dwarfs, and neutron stars: The physics of compact objects}}}\ (\bibinfo
  {publisher} {John Wiley \& Sons},\ \bibinfo {year} {2008})\BibitemShut
  {NoStop}%
\bibitem [{\citenamefont {Perli{\'n}ska}\ \emph {et~al.}(2004)\citenamefont
  {Perli{\'n}ska}, \citenamefont {Rohozi{\'n}ski}, \citenamefont
  {Dobaczewski},\ and\ \citenamefont {Nazarewicz}}]{perlinska2004local}%
  \BibitemOpen
  \bibfield  {author} {\bibinfo {author} {\bibfnamefont {E.}~\bibnamefont
  {Perli{\'n}ska}}, \bibinfo {author} {\bibfnamefont {S.}~\bibnamefont
  {Rohozi{\'n}ski}}, \bibinfo {author} {\bibfnamefont {J.}~\bibnamefont
  {Dobaczewski}},\ and\ \bibinfo {author} {\bibfnamefont {W.}~\bibnamefont
  {Nazarewicz}},\ }\bibfield  {title} {\bibinfo {title} {Local density
  approximation for proton-neutron pairing correlations: Formalism},\
  }\href@noop {} {\bibfield  {journal} {\bibinfo  {journal} {Physical Review
  C}\ }\textbf {\bibinfo {volume} {69}},\ \bibinfo {pages} {014316} (\bibinfo
  {year} {2004})}\BibitemShut {NoStop}%
\bibitem [{\citenamefont {Becker}\ \emph {et~al.}(2017)\citenamefont {Becker},
  \citenamefont {Davesne}, \citenamefont {Meyer}, \citenamefont {Navarro},\
  and\ \citenamefont {Pastore}}]{becker2017solution}%
  \BibitemOpen
  \bibfield  {author} {\bibinfo {author} {\bibfnamefont {P.}~\bibnamefont
  {Becker}}, \bibinfo {author} {\bibfnamefont {D.}~\bibnamefont {Davesne}},
  \bibinfo {author} {\bibfnamefont {J.}~\bibnamefont {Meyer}}, \bibinfo
  {author} {\bibfnamefont {J.}~\bibnamefont {Navarro}},\ and\ \bibinfo {author}
  {\bibfnamefont {A.}~\bibnamefont {Pastore}},\ }\bibfield  {title} {\bibinfo
  {title} {Solution of hartree-fock-bogoliubov equations and fitting procedure
  using the n2lo skyrme pseudopotential in spherical symmetry},\ }\href@noop {}
  {\bibfield  {journal} {\bibinfo  {journal} {Physical Review C}\ }\textbf
  {\bibinfo {volume} {96}},\ \bibinfo {pages} {044330} (\bibinfo {year}
  {2017})}\BibitemShut {NoStop}%
\bibitem [{\citenamefont {Ring}\ and\ \citenamefont
  {Schuck}(2004)}]{ring2004nuclear}%
  \BibitemOpen
  \bibfield  {author} {\bibinfo {author} {\bibfnamefont {P.}~\bibnamefont
  {Ring}}\ and\ \bibinfo {author} {\bibfnamefont {P.}~\bibnamefont {Schuck}},\
  }\href@noop {} {\emph {\bibinfo {title} {The nuclear many-body problem}}}\
  (\bibinfo  {publisher} {Springer Science \& Business Media},\ \bibinfo {year}
  {2004})\BibitemShut {NoStop}%
\bibitem [{\citenamefont {Grammaticos}\ and\ \citenamefont
  {Voros}(1979)}]{Grammaticos1979}%
  \BibitemOpen
  \bibfield  {author} {\bibinfo {author} {\bibfnamefont {B.}~\bibnamefont
  {Grammaticos}}\ and\ \bibinfo {author} {\bibfnamefont {A.}~\bibnamefont
  {Voros}},\ }\bibfield  {title} {\bibinfo {title} {{Semiclassical
  approximations for nuclear hamiltonians. I. Spin-Independent Potentials}},\
  }\href
  {http://www.osti.gov/energycitations/product.biblio.jsp?osti\_id=5359248}
  {\bibfield  {journal} {\bibinfo  {journal} {Annals of Physics}\ }\textbf
  {\bibinfo {volume} {123}},\ \bibinfo {pages} {359} (\bibinfo {year}
  {1979})}\BibitemShut {NoStop}%
\bibitem [{\citenamefont {Bartel}\ and\ \citenamefont
  {Bencheikh}(2002)}]{bartel2002nuclear}%
  \BibitemOpen
  \bibfield  {author} {\bibinfo {author} {\bibfnamefont {J.}~\bibnamefont
  {Bartel}}\ and\ \bibinfo {author} {\bibfnamefont {K.}~\bibnamefont
  {Bencheikh}},\ }\bibfield  {title} {\bibinfo {title} {Nuclear mean fields
  through self-consistent semiclassical calculations},\ }\href@noop {}
  {\bibfield  {journal} {\bibinfo  {journal} {The European Physical Journal
  A-Hadrons and Nuclei}\ }\textbf {\bibinfo {volume} {14}},\ \bibinfo {pages}
  {179} (\bibinfo {year} {2002})}\BibitemShut {NoStop}%
\bibitem [{\citenamefont {Sandulescu}(2004)}]{sandulescu2004nuclear}%
  \BibitemOpen
  \bibfield  {author} {\bibinfo {author} {\bibfnamefont {N.}~\bibnamefont
  {Sandulescu}},\ }\bibfield  {title} {\bibinfo {title} {Nuclear superfluidity
  and specific heat in the inner crust of neutron stars},\ }\href@noop {}
  {\bibfield  {journal} {\bibinfo  {journal} {Physical Review C}\ }\textbf
  {\bibinfo {volume} {70}},\ \bibinfo {pages} {025801} (\bibinfo {year}
  {2004})}\BibitemShut {NoStop}%
\bibitem [{\citenamefont {Watanabe}\ and\ \citenamefont
  {Pethick}(2017)}]{watanabe2017superfluid}%
  \BibitemOpen
  \bibfield  {author} {\bibinfo {author} {\bibfnamefont {G.}~\bibnamefont
  {Watanabe}}\ and\ \bibinfo {author} {\bibfnamefont {C.~J.}\ \bibnamefont
  {Pethick}},\ }\bibfield  {title} {\bibinfo {title} {Superfluid density of
  neutrons in the inner crust of neutron stars: new life for pulsar glitch
  models},\ }\href@noop {} {\bibfield  {journal} {\bibinfo  {journal} {Physical
  Review Letters}\ }\textbf {\bibinfo {volume} {119}},\ \bibinfo {pages}
  {062701} (\bibinfo {year} {2017})}\BibitemShut {NoStop}%
\bibitem [{\citenamefont {Baldo}\ \emph {et~al.}(2005)\citenamefont {Baldo},
  \citenamefont {Lombardo}, \citenamefont {Saperstein},\ and\ \citenamefont
  {Tolokonnikov}}]{baldo2005role}%
  \BibitemOpen
  \bibfield  {author} {\bibinfo {author} {\bibfnamefont {M.}~\bibnamefont
  {Baldo}}, \bibinfo {author} {\bibfnamefont {U.}~\bibnamefont {Lombardo}},
  \bibinfo {author} {\bibfnamefont {E.}~\bibnamefont {Saperstein}},\ and\
  \bibinfo {author} {\bibfnamefont {S.}~\bibnamefont {Tolokonnikov}},\
  }\bibfield  {title} {\bibinfo {title} {The role of superfluidity in the
  structure of the neutron star inner crust},\ }\href@noop {} {\bibfield
  {journal} {\bibinfo  {journal} {Nuclear Physics A}\ }\textbf {\bibinfo
  {volume} {750}},\ \bibinfo {pages} {409} (\bibinfo {year}
  {2005})}\BibitemShut {NoStop}%
\bibitem [{\citenamefont {Bertsch}\ and\ \citenamefont
  {Esbensen}(1991)}]{bertsch1991pair}%
  \BibitemOpen
  \bibfield  {author} {\bibinfo {author} {\bibfnamefont {G.}~\bibnamefont
  {Bertsch}}\ and\ \bibinfo {author} {\bibfnamefont {H.}~\bibnamefont
  {Esbensen}},\ }\bibfield  {title} {\bibinfo {title} {Pair correlations near
  the neutron drip line},\ }\href@noop {} {\bibfield  {journal} {\bibinfo
  {journal} {Annals of Physics}\ }\textbf {\bibinfo {volume} {209}},\ \bibinfo
  {pages} {327} (\bibinfo {year} {1991})}\BibitemShut {NoStop}%
\bibitem [{\citenamefont {Gandolfi}\ \emph {et~al.}(2008)\citenamefont
  {Gandolfi}, \citenamefont {Illarionov}, \citenamefont {Fantoni},
  \citenamefont {Pederiva},\ and\ \citenamefont
  {Schmidt}}]{gandolfi2008equation}%
  \BibitemOpen
  \bibfield  {author} {\bibinfo {author} {\bibfnamefont {S.}~\bibnamefont
  {Gandolfi}}, \bibinfo {author} {\bibfnamefont {A.~Y.}\ \bibnamefont
  {Illarionov}}, \bibinfo {author} {\bibfnamefont {S.}~\bibnamefont {Fantoni}},
  \bibinfo {author} {\bibfnamefont {F.}~\bibnamefont {Pederiva}},\ and\
  \bibinfo {author} {\bibfnamefont {K.}~\bibnamefont {Schmidt}},\ }\bibfield
  {title} {\bibinfo {title} {Equation of state of superfluid neutron matter and
  the calculation of the s 0 1 pairing gap},\ }\href@noop {} {\bibfield
  {journal} {\bibinfo  {journal} {Physical review letters}\ }\textbf {\bibinfo
  {volume} {101}},\ \bibinfo {pages} {132501} (\bibinfo {year}
  {2008})}\BibitemShut {NoStop}%
\bibitem [{\citenamefont {Bulgac}\ and\ \citenamefont
  {Yu}(2002)}]{bulgac2002renormalization}%
  \BibitemOpen
  \bibfield  {author} {\bibinfo {author} {\bibfnamefont {A.}~\bibnamefont
  {Bulgac}}\ and\ \bibinfo {author} {\bibfnamefont {Y.}~\bibnamefont {Yu}},\
  }\bibfield  {title} {\bibinfo {title} {Renormalization of the
  hartree-fock-bogoliubov equations in the case of a zero range pairing
  interaction},\ }\href@noop {} {\bibfield  {journal} {\bibinfo  {journal}
  {Physical review letters}\ }\textbf {\bibinfo {volume} {88}},\ \bibinfo
  {pages} {042504} (\bibinfo {year} {2002})}\BibitemShut {NoStop}%
\bibitem [{\citenamefont {Goriely}\ \emph {et~al.}(2013)\citenamefont
  {Goriely}, \citenamefont {Chamel},\ and\ \citenamefont
  {Pearson}}]{goriely2013further}%
  \BibitemOpen
  \bibfield  {author} {\bibinfo {author} {\bibfnamefont {S.}~\bibnamefont
  {Goriely}}, \bibinfo {author} {\bibfnamefont {N.}~\bibnamefont {Chamel}},\
  and\ \bibinfo {author} {\bibfnamefont {J.}~\bibnamefont {Pearson}},\
  }\bibfield  {title} {\bibinfo {title} {Further explorations of
  skyrme-hartree-fock-bogoliubov mass formulas. xiii. the 2012 atomic mass
  evaluation and the symmetry coefficient},\ }\href@noop {} {\bibfield
  {journal} {\bibinfo  {journal} {Physical Review C}\ }\textbf {\bibinfo
  {volume} {88}},\ \bibinfo {pages} {024308} (\bibinfo {year}
  {2013})}\BibitemShut {NoStop}%
\bibitem [{\citenamefont {Piekarewicz}\ and\ \citenamefont
  {Centelles}(2009)}]{piekarewicz2009incompressibility}%
  \BibitemOpen
  \bibfield  {author} {\bibinfo {author} {\bibfnamefont {J.}~\bibnamefont
  {Piekarewicz}}\ and\ \bibinfo {author} {\bibfnamefont {M.}~\bibnamefont
  {Centelles}},\ }\bibfield  {title} {\bibinfo {title} {Incompressibility of
  neutron-rich matter},\ }\href@noop {} {\bibfield  {journal} {\bibinfo
  {journal} {Physical Review C}\ }\textbf {\bibinfo {volume} {79}},\ \bibinfo
  {pages} {054311} (\bibinfo {year} {2009})}\BibitemShut {NoStop}%
\bibitem [{\citenamefont {Tsang}\ \emph
  {et~al.}(2012{\natexlab{b}})\citenamefont {Tsang}, \citenamefont {Stone},
  \citenamefont {Camera}, \citenamefont {Danielewicz}, \citenamefont
  {Gandolfi}, \citenamefont {Hebeler}, \citenamefont {Horowitz}, \citenamefont
  {Lee}, \citenamefont {Lynch}, \citenamefont {Kohley} \emph
  {et~al.}}]{tsang2012constraints}%
  \BibitemOpen
  \bibfield  {author} {\bibinfo {author} {\bibfnamefont {M.}~\bibnamefont
  {Tsang}}, \bibinfo {author} {\bibfnamefont {J.}~\bibnamefont {Stone}},
  \bibinfo {author} {\bibfnamefont {F.}~\bibnamefont {Camera}}, \bibinfo
  {author} {\bibfnamefont {P.}~\bibnamefont {Danielewicz}}, \bibinfo {author}
  {\bibfnamefont {S.}~\bibnamefont {Gandolfi}}, \bibinfo {author}
  {\bibfnamefont {K.}~\bibnamefont {Hebeler}}, \bibinfo {author} {\bibfnamefont
  {C.~J.}\ \bibnamefont {Horowitz}}, \bibinfo {author} {\bibfnamefont
  {J.}~\bibnamefont {Lee}}, \bibinfo {author} {\bibfnamefont {W.~G.}\
  \bibnamefont {Lynch}}, \bibinfo {author} {\bibfnamefont {Z.}~\bibnamefont
  {Kohley}}, \emph {et~al.},\ }\bibfield  {title} {\bibinfo {title}
  {Constraints on the symmetry energy and neutron skins from experiments and
  theory},\ }\href@noop {} {\bibfield  {journal} {\bibinfo  {journal} {Physical
  Review C}\ }\textbf {\bibinfo {volume} {86}},\ \bibinfo {pages} {015803}
  (\bibinfo {year} {2012}{\natexlab{b}})}\BibitemShut {NoStop}%
\bibitem [{\citenamefont {Chen}\ \emph {et~al.}(2010)\citenamefont {Chen},
  \citenamefont {Ko}, \citenamefont {Li},\ and\ \citenamefont
  {Xu}}]{chen2010density}%
  \BibitemOpen
  \bibfield  {author} {\bibinfo {author} {\bibfnamefont {L.-W.}\ \bibnamefont
  {Chen}}, \bibinfo {author} {\bibfnamefont {C.~M.}\ \bibnamefont {Ko}},
  \bibinfo {author} {\bibfnamefont {B.-A.}\ \bibnamefont {Li}},\ and\ \bibinfo
  {author} {\bibfnamefont {J.}~\bibnamefont {Xu}},\ }\bibfield  {title}
  {\bibinfo {title} {Density slope of the nuclear symmetry energy from the
  neutron skin thickness of heavy nuclei},\ }\href@noop {} {\bibfield
  {journal} {\bibinfo  {journal} {Physical Review C}\ }\textbf {\bibinfo
  {volume} {82}},\ \bibinfo {pages} {024321} (\bibinfo {year}
  {2010})}\BibitemShut {NoStop}%
\bibitem [{\citenamefont {Vidana}\ \emph {et~al.}(2009)\citenamefont {Vidana},
  \citenamefont {Provid{\^e}ncia}, \citenamefont {Polls},\ and\ \citenamefont
  {Rios}}]{vidana2009density}%
  \BibitemOpen
  \bibfield  {author} {\bibinfo {author} {\bibfnamefont {I.}~\bibnamefont
  {Vidana}}, \bibinfo {author} {\bibfnamefont {C.}~\bibnamefont
  {Provid{\^e}ncia}}, \bibinfo {author} {\bibfnamefont {A.}~\bibnamefont
  {Polls}},\ and\ \bibinfo {author} {\bibfnamefont {A.}~\bibnamefont {Rios}},\
  }\bibfield  {title} {\bibinfo {title} {Density dependence of the nuclear
  symmetry energy: A microscopic perspective},\ }\href@noop {} {\bibfield
  {journal} {\bibinfo  {journal} {Physical Review C}\ }\textbf {\bibinfo
  {volume} {80}},\ \bibinfo {pages} {045806} (\bibinfo {year}
  {2009})}\BibitemShut {NoStop}%
\bibitem [{\citenamefont {Roca-Maza}\ \emph {et~al.}(2011)\citenamefont
  {Roca-Maza}, \citenamefont {Centelles}, \citenamefont {Vinas},\ and\
  \citenamefont {Warda}}]{roca2011neutron}%
  \BibitemOpen
  \bibfield  {author} {\bibinfo {author} {\bibfnamefont {X.}~\bibnamefont
  {Roca-Maza}}, \bibinfo {author} {\bibfnamefont {M.}~\bibnamefont
  {Centelles}}, \bibinfo {author} {\bibfnamefont {X.}~\bibnamefont {Vinas}},\
  and\ \bibinfo {author} {\bibfnamefont {M.}~\bibnamefont {Warda}},\ }\bibfield
   {title} {\bibinfo {title} {Neutron skin of pb 208, nuclear symmetry energy,
  and the parity radius experiment},\ }\href@noop {} {\bibfield  {journal}
  {\bibinfo  {journal} {Physical review letters}\ }\textbf {\bibinfo {volume}
  {106}},\ \bibinfo {pages} {252501} (\bibinfo {year} {2011})}\BibitemShut
  {NoStop}%
\bibitem [{\citenamefont {Li}\ \emph {et~al.}(2014)\citenamefont {Li},
  \citenamefont {Ramos}, \citenamefont {Verde},\ and\ \citenamefont
  {Vida{\~n}a}}]{liTopicalIssueNuclear2014}%
  \BibitemOpen
  \bibfield  {author} {\bibinfo {author} {\bibfnamefont {B.-A.}\ \bibnamefont
  {Li}}, \bibinfo {author} {\bibfnamefont {{\`A}.}~\bibnamefont {Ramos}},
  \bibinfo {author} {\bibfnamefont {G.}~\bibnamefont {Verde}},\ and\ \bibinfo
  {author} {\bibfnamefont {I.}~\bibnamefont {Vida{\~n}a}},\ }\bibfield  {title}
  {\bibinfo {title} {Topical issue on nuclear symmetry energy},\ }\href
  {https://doi.org/10.1140/epja/i2014-14009-x} {\bibfield  {journal} {\bibinfo
  {journal} {The European Physical Journal A}\ }\textbf {\bibinfo {volume}
  {50}},\ \bibinfo {pages} {9} (\bibinfo {year} {2014})}\BibitemShut {NoStop}%
\bibitem [{\citenamefont {Drischler}\ \emph {et~al.}(2019)\citenamefont
  {Drischler}, \citenamefont {Hebeler},\ and\ \citenamefont
  {Schwenk}}]{drischlerChiralInteractionsNexttoNexttoNexttoLeading2019}%
  \BibitemOpen
  \bibfield  {author} {\bibinfo {author} {\bibfnamefont {C.}~\bibnamefont
  {Drischler}}, \bibinfo {author} {\bibfnamefont {K.}~\bibnamefont {Hebeler}},\
  and\ \bibinfo {author} {\bibfnamefont {A.}~\bibnamefont {Schwenk}},\
  }\bibfield  {title} {\bibinfo {title} {Chiral {{Interactions}} up to
  {{Next}}-to-{{Next}}-to-{{Next}}-to-{{Leading Order}} and {{Nuclear
  Saturation}}},\ }\href {https://doi.org/10.1103/PhysRevLett.122.042501}
  {\bibfield  {journal} {\bibinfo  {journal} {Physical Review Letters}\
  }\textbf {\bibinfo {volume} {122}},\ \bibinfo {pages} {042501} (\bibinfo
  {year} {2019})}\BibitemShut {NoStop}%
\bibitem [{\citenamefont {Drischler}\ \emph {et~al.}(2020)\citenamefont
  {Drischler}, \citenamefont {Furnstahl}, \citenamefont {Melendez},\ and\
  \citenamefont {Phillips}}]{drischlerHowWellWe2020}%
  \BibitemOpen
  \bibfield  {author} {\bibinfo {author} {\bibfnamefont {C.}~\bibnamefont
  {Drischler}}, \bibinfo {author} {\bibfnamefont {R.~J.}\ \bibnamefont
  {Furnstahl}}, \bibinfo {author} {\bibfnamefont {J.~A.}\ \bibnamefont
  {Melendez}},\ and\ \bibinfo {author} {\bibfnamefont {D.~R.}\ \bibnamefont
  {Phillips}},\ }\bibfield  {title} {\bibinfo {title} {How {{Well Do We Know}}
  the {{Neutron}}-{{Matter Equation}} of {{State}} at the {{Densities Inside
  Neutron Stars}}? {{A Bayesian Approach}} with {{Correlated Uncertainties}}},\
  }\href {https://doi.org/10.1103/PhysRevLett.125.202702} {\bibfield  {journal}
  {\bibinfo  {journal} {Physical Review Letters}\ }\textbf {\bibinfo {volume}
  {125}},\ \bibinfo {pages} {202702} (\bibinfo {year} {2020})}\BibitemShut
  {NoStop}%
\bibitem [{\citenamefont {Agrawal}\ \emph {et~al.}(2005)\citenamefont
  {Agrawal}, \citenamefont {Shlomo},\ and\ \citenamefont
  {Au}}]{agrawal2005determination}%
  \BibitemOpen
  \bibfield  {author} {\bibinfo {author} {\bibfnamefont {B.}~\bibnamefont
  {Agrawal}}, \bibinfo {author} {\bibfnamefont {S.}~\bibnamefont {Shlomo}},\
  and\ \bibinfo {author} {\bibfnamefont {V.~K.}\ \bibnamefont {Au}},\
  }\bibfield  {title} {\bibinfo {title} {Determination of the parameters of a
  skyrme type effective interaction using the simulated annealing approach},\
  }\href@noop {} {\bibfield  {journal} {\bibinfo  {journal} {Physical Review
  C}\ }\textbf {\bibinfo {volume} {72}},\ \bibinfo {pages} {014310} (\bibinfo
  {year} {2005})}\BibitemShut {NoStop}%
\bibitem [{\citenamefont {Cao}\ \emph {et~al.}(2006)\citenamefont {Cao},
  \citenamefont {Lombardo}, \citenamefont {Shen},\ and\ \citenamefont
  {Van~Giai}}]{cao2006brueckner}%
  \BibitemOpen
  \bibfield  {author} {\bibinfo {author} {\bibfnamefont {L.}~\bibnamefont
  {Cao}}, \bibinfo {author} {\bibfnamefont {U.}~\bibnamefont {Lombardo}},
  \bibinfo {author} {\bibfnamefont {C.}~\bibnamefont {Shen}},\ and\ \bibinfo
  {author} {\bibfnamefont {N.}~\bibnamefont {Van~Giai}},\ }\bibfield  {title}
  {\bibinfo {title} {From brueckner approach to skyrme-type energy density
  functional},\ }\href@noop {} {\bibfield  {journal} {\bibinfo  {journal}
  {Physical Review C}\ }\textbf {\bibinfo {volume} {73}},\ \bibinfo {pages}
  {014313} (\bibinfo {year} {2006})}\BibitemShut {NoStop}%
\bibitem [{\citenamefont {Stevenson}\ \emph {et~al.}(2013)\citenamefont
  {Stevenson}, \citenamefont {Goddard}, \citenamefont {Stone},\ and\
  \citenamefont {Dutra}}]{stevenson2013skyrme}%
  \BibitemOpen
  \bibfield  {author} {\bibinfo {author} {\bibfnamefont {P.}~\bibnamefont
  {Stevenson}}, \bibinfo {author} {\bibfnamefont {P.}~\bibnamefont {Goddard}},
  \bibinfo {author} {\bibfnamefont {J.}~\bibnamefont {Stone}},\ and\ \bibinfo
  {author} {\bibfnamefont {M.}~\bibnamefont {Dutra}},\ }\bibfield  {title}
  {\bibinfo {title} {Do skyrme forces that fit nuclear matter work well in
  finite nuclei?},\ }in\ \href@noop {} {\emph {\bibinfo {booktitle} {AIP
  Conference Proceedings}}},\ Vol.\ \bibinfo {volume} {1529}\ (\bibinfo
  {organization} {American Institute of Physics},\ \bibinfo {year} {2013})\
  pp.\ \bibinfo {pages} {262--268}\BibitemShut {NoStop}%
\bibitem [{\citenamefont {Steiner}\ \emph {et~al.}(2005)\citenamefont
  {Steiner}, \citenamefont {Prakash}, \citenamefont {Lattimer},\ and\
  \citenamefont {Ellis}}]{steiner2005isospin}%
  \BibitemOpen
  \bibfield  {author} {\bibinfo {author} {\bibfnamefont {A.~W.}\ \bibnamefont
  {Steiner}}, \bibinfo {author} {\bibfnamefont {M.}~\bibnamefont {Prakash}},
  \bibinfo {author} {\bibfnamefont {J.~M.}\ \bibnamefont {Lattimer}},\ and\
  \bibinfo {author} {\bibfnamefont {P.~J.}\ \bibnamefont {Ellis}},\ }\bibfield
  {title} {\bibinfo {title} {Isospin asymmetry in nuclei and neutron stars},\
  }\href@noop {} {\bibfield  {journal} {\bibinfo  {journal} {Physics reports}\
  }\textbf {\bibinfo {volume} {411}},\ \bibinfo {pages} {325} (\bibinfo {year}
  {2005})}\BibitemShut {NoStop}%
\bibitem [{\citenamefont {Vautherin}\ and\ \citenamefont
  {Brink}(1972)}]{vautherin1972hartree}%
  \BibitemOpen
  \bibfield  {author} {\bibinfo {author} {\bibfnamefont {D.}~\bibnamefont
  {Vautherin}}\ and\ \bibinfo {author} {\bibfnamefont {D.~t.}\ \bibnamefont
  {Brink}},\ }\bibfield  {title} {\bibinfo {title} {Hartree-fock calculations
  with skyrme's interaction. i. spherical nuclei},\ }\href@noop {} {\bibfield
  {journal} {\bibinfo  {journal} {Physical Review C}\ }\textbf {\bibinfo
  {volume} {5}},\ \bibinfo {pages} {626} (\bibinfo {year} {1972})}\BibitemShut
  {NoStop}%
\bibitem [{\citenamefont {Beiner}\ \emph {et~al.}(1975)\citenamefont {Beiner},
  \citenamefont {Flocard}, \citenamefont {Van~Giai},\ and\ \citenamefont
  {Quentin}}]{beiner1975nuclear}%
  \BibitemOpen
  \bibfield  {author} {\bibinfo {author} {\bibfnamefont {M.}~\bibnamefont
  {Beiner}}, \bibinfo {author} {\bibfnamefont {H.}~\bibnamefont {Flocard}},
  \bibinfo {author} {\bibfnamefont {N.}~\bibnamefont {Van~Giai}},\ and\
  \bibinfo {author} {\bibfnamefont {P.}~\bibnamefont {Quentin}},\ }\bibfield
  {title} {\bibinfo {title} {Nuclear ground-state properties and
  self-consistent calculations with the skyrme interaction:(i). spherical
  description},\ }\href@noop {} {\bibfield  {journal} {\bibinfo  {journal}
  {Nuclear Physics A}\ }\textbf {\bibinfo {volume} {238}},\ \bibinfo {pages}
  {29} (\bibinfo {year} {1975})}\BibitemShut {NoStop}%
\bibitem [{\citenamefont {Rashdan}(2000)}]{rashdan2000skyrme}%
  \BibitemOpen
  \bibfield  {author} {\bibinfo {author} {\bibfnamefont {M.}~\bibnamefont
  {Rashdan}},\ }\bibfield  {title} {\bibinfo {title} {A skyrme parametrization
  based on nuclear matter bhf calculations},\ }\href@noop {} {\bibfield
  {journal} {\bibinfo  {journal} {Modern Physics Letters A}\ }\textbf {\bibinfo
  {volume} {15}},\ \bibinfo {pages} {1287} (\bibinfo {year}
  {2000})}\BibitemShut {NoStop}%
\bibitem [{\citenamefont {K{\"o}hler}(1976)}]{kohler1976skyrme}%
  \BibitemOpen
  \bibfield  {author} {\bibinfo {author} {\bibfnamefont {H.}~\bibnamefont
  {K{\"o}hler}},\ }\bibfield  {title} {\bibinfo {title} {Skyrme force and the
  mass formula},\ }\href@noop {} {\bibfield  {journal} {\bibinfo  {journal}
  {Nuclear Physics A}\ }\textbf {\bibinfo {volume} {258}},\ \bibinfo {pages}
  {301} (\bibinfo {year} {1976})}\BibitemShut {NoStop}%
\bibitem [{\citenamefont {Chabanat}\ \emph {et~al.}(1998)\citenamefont
  {Chabanat}, \citenamefont {Bonche}, \citenamefont {Haensel}, \citenamefont
  {Meyer},\ and\ \citenamefont {Schaeffer}}]{chabanat1998skyrme}%
  \BibitemOpen
  \bibfield  {author} {\bibinfo {author} {\bibfnamefont {E.}~\bibnamefont
  {Chabanat}}, \bibinfo {author} {\bibfnamefont {P.}~\bibnamefont {Bonche}},
  \bibinfo {author} {\bibfnamefont {P.}~\bibnamefont {Haensel}}, \bibinfo
  {author} {\bibfnamefont {J.}~\bibnamefont {Meyer}},\ and\ \bibinfo {author}
  {\bibfnamefont {R.}~\bibnamefont {Schaeffer}},\ }\bibfield  {title} {\bibinfo
  {title} {A skyrme parametrization from subnuclear to neutron star densities
  part ii. nuclei far from stabilities},\ }\href@noop {} {\bibfield  {journal}
  {\bibinfo  {journal} {Nuclear Physics A}\ }\textbf {\bibinfo {volume}
  {635}},\ \bibinfo {pages} {231} (\bibinfo {year} {1998})}\BibitemShut
  {NoStop}%
\bibitem [{\citenamefont {Guichon}\ \emph {et~al.}(2006)\citenamefont
  {Guichon}, \citenamefont {Matevosyan}, \citenamefont {Sandulescu},\ and\
  \citenamefont {Thomas}}]{guichon2006physical}%
  \BibitemOpen
  \bibfield  {author} {\bibinfo {author} {\bibfnamefont {P.~A.}\ \bibnamefont
  {Guichon}}, \bibinfo {author} {\bibfnamefont {H.~H.}\ \bibnamefont
  {Matevosyan}}, \bibinfo {author} {\bibfnamefont {N.}~\bibnamefont
  {Sandulescu}},\ and\ \bibinfo {author} {\bibfnamefont {A.~W.}\ \bibnamefont
  {Thomas}},\ }\bibfield  {title} {\bibinfo {title} {Physical origin of density
  dependent forces of skyrme type within the quark meson coupling model},\
  }\href@noop {} {\bibfield  {journal} {\bibinfo  {journal} {Nuclear Physics
  A}\ }\textbf {\bibinfo {volume} {772}},\ \bibinfo {pages} {1} (\bibinfo
  {year} {2006})}\BibitemShut {NoStop}%
\bibitem [{\citenamefont {Krivine}\ \emph {et~al.}(1980)\citenamefont
  {Krivine}, \citenamefont {Treiner},\ and\ \citenamefont
  {Bohigas}}]{krivine1980derivation}%
  \BibitemOpen
  \bibfield  {author} {\bibinfo {author} {\bibfnamefont {H.}~\bibnamefont
  {Krivine}}, \bibinfo {author} {\bibfnamefont {J.}~\bibnamefont {Treiner}},\
  and\ \bibinfo {author} {\bibfnamefont {O.}~\bibnamefont {Bohigas}},\
  }\bibfield  {title} {\bibinfo {title} {Derivation of a fluid-dynamical
  lagrangian and electric giant resonances},\ }\href@noop {} {\bibfield
  {journal} {\bibinfo  {journal} {Nuclear Physics A}\ }\textbf {\bibinfo
  {volume} {336}},\ \bibinfo {pages} {155} (\bibinfo {year}
  {1980})}\BibitemShut {NoStop}%
\bibitem [{\citenamefont {Margueron}\ \emph {et~al.}(2002)\citenamefont
  {Margueron}, \citenamefont {Navarro},\ and\ \citenamefont
  {Van~Giai}}]{margueron2002instabilities}%
  \BibitemOpen
  \bibfield  {author} {\bibinfo {author} {\bibfnamefont {J.}~\bibnamefont
  {Margueron}}, \bibinfo {author} {\bibfnamefont {J.}~\bibnamefont {Navarro}},\
  and\ \bibinfo {author} {\bibfnamefont {N.}~\bibnamefont {Van~Giai}},\
  }\bibfield  {title} {\bibinfo {title} {Instabilities of infinite matter with
  effective skyrme-type interactions},\ }\href@noop {} {\bibfield  {journal}
  {\bibinfo  {journal} {Physical Review C}\ }\textbf {\bibinfo {volume} {66}},\
  \bibinfo {pages} {014303} (\bibinfo {year} {2002})}\BibitemShut {NoStop}%
\bibitem [{\citenamefont {Tews}\ \emph {et~al.}(2013)\citenamefont {Tews},
  \citenamefont {Kr\"uger}, \citenamefont {Hebeler},\ and\ \citenamefont
  {Schwenk}}]{Tews}%
  \BibitemOpen
  \bibfield  {author} {\bibinfo {author} {\bibfnamefont {I.}~\bibnamefont
  {Tews}}, \bibinfo {author} {\bibfnamefont {T.}~\bibnamefont {Kr\"uger}},
  \bibinfo {author} {\bibfnamefont {K.}~\bibnamefont {Hebeler}},\ and\ \bibinfo
  {author} {\bibfnamefont {A.}~\bibnamefont {Schwenk}},\ }\bibfield  {title}
  {\bibinfo {title} {Neutron matter at next-to-next-to-next-to-leading order in
  chiral effective field theory},\ }\href
  {https://doi.org/10.1103/PhysRevLett.110.032504} {\bibfield  {journal}
  {\bibinfo  {journal} {Phys. Rev. Lett.}\ }\textbf {\bibinfo {volume} {110}},\
  \bibinfo {pages} {032504} (\bibinfo {year} {2013})}\BibitemShut {NoStop}%
\bibitem [{\citenamefont {Davesne}\ \emph {et~al.}(2016)\citenamefont
  {Davesne}, \citenamefont {Pastore},\ and\ \citenamefont
  {Navarro}}]{davesne2016extended}%
  \BibitemOpen
  \bibfield  {author} {\bibinfo {author} {\bibfnamefont {D.}~\bibnamefont
  {Davesne}}, \bibinfo {author} {\bibfnamefont {A.}~\bibnamefont {Pastore}},\
  and\ \bibinfo {author} {\bibfnamefont {J.}~\bibnamefont {Navarro}},\
  }\bibfield  {title} {\bibinfo {title} {Extended skyrme equation of state in
  asymmetric nuclear matter},\ }\href@noop {} {\bibfield  {journal} {\bibinfo
  {journal} {Astronomy \& Astrophysics}\ }\textbf {\bibinfo {volume} {585}},\
  \bibinfo {pages} {A83} (\bibinfo {year} {2016})}\BibitemShut {NoStop}%
\bibitem [{\citenamefont {Iwamoto}\ and\ \citenamefont
  {Pethick}(1982)}]{iwamoto1982effects}%
  \BibitemOpen
  \bibfield  {author} {\bibinfo {author} {\bibfnamefont {N.}~\bibnamefont
  {Iwamoto}}\ and\ \bibinfo {author} {\bibfnamefont {C.}~\bibnamefont
  {Pethick}},\ }\bibfield  {title} {\bibinfo {title} {Effects of
  nucleon-nucleon interactions on scattering of neutrinos in neutron matter},\
  }\href@noop {} {\bibfield  {journal} {\bibinfo  {journal} {Physical Review
  D}\ }\textbf {\bibinfo {volume} {25}},\ \bibinfo {pages} {313} (\bibinfo
  {year} {1982})}\BibitemShut {NoStop}%
\bibitem [{\citenamefont {Li}\ and\ \citenamefont
  {Schulze}(2008)}]{liNeutronStarStructure2008}%
  \BibitemOpen
  \bibfield  {author} {\bibinfo {author} {\bibfnamefont {Z.~H.}\ \bibnamefont
  {Li}}\ and\ \bibinfo {author} {\bibfnamefont {H.-J.}\ \bibnamefont
  {Schulze}},\ }\bibfield  {title} {\bibinfo {title} {Neutron star structure
  with modern nucleonic three-body forces},\ }\href
  {https://doi.org/10.1103/PhysRevC.78.028801} {\bibfield  {journal} {\bibinfo
  {journal} {Physical Review C}\ }\textbf {\bibinfo {volume} {78}},\ \bibinfo
  {pages} {028801} (\bibinfo {year} {2008})}\BibitemShut {NoStop}%
\bibitem [{\citenamefont {Zdunik}\ \emph {et~al.}(2017)\citenamefont {Zdunik},
  \citenamefont {Fortin},\ and\ \citenamefont {Haensel}}]{zdunik2017neutron}%
  \BibitemOpen
  \bibfield  {author} {\bibinfo {author} {\bibfnamefont {J.}~\bibnamefont
  {Zdunik}}, \bibinfo {author} {\bibfnamefont {M.}~\bibnamefont {Fortin}},\
  and\ \bibinfo {author} {\bibfnamefont {P.}~\bibnamefont {Haensel}},\
  }\bibfield  {title} {\bibinfo {title} {Neutron star properties and the
  equation of state for the core},\ }\href@noop {} {\bibfield  {journal}
  {\bibinfo  {journal} {Astronomy \& Astrophysics}\ }\textbf {\bibinfo {volume}
  {599}},\ \bibinfo {pages} {A119} (\bibinfo {year} {2017})}\BibitemShut
  {NoStop}%
\end{thebibliography}%

\end{document}